\documentclass[vecphys]{svmult}

\usepackage{makeidx}         
\usepackage{graphicx}        
\usepackage{multicol}        
\usepackage{amsmath}

\makeindex             

\begin{document}

\title*{
Current rectification, switching, polarons, and defects in molecular
electronic devices
  }
\titlerunning{Electron transport, switching, polarons in molecular devices
}

\author{A.M. Bratkovsky}

\institute{ Hewlett-Packard Laboratories, 1501 Page Mill Road, Palo
Alto, California 94304 }
%
%
\maketitle

Devices for nano- and molecular size electronics are currently a
focus of research aimed at an efficient current rectification and
switching.\footnote{Chapter from: {\em Polarons in Advanced
Materials}, edited by A. S. Alexandrov (Canopus/Springer, Bristol,
2007).}
 A few generic molecular scale devices are reviewed here
on the basis of first-principles and model approaches. Current
rectification by (ballistic) molecular quantum dots can produce the
rectification ratio $\leq 100$. Current switching due to
conformational changes in the molecules is slow, on the order of a
few kHz. Fast switching ($\sim $1~THz) may be achieved, at least in
principle, in a degenerate molecular quantum dot with strong
coupling of electrons with vibrational excitations. We show that the
mean-field approach fails to properly describe \emph{intrinsic}
molecular switching and present an exact solution to the problem.
Defects in molecular films result in spurious peaks in conductance,
apparent negative differential resistance, and may also lead to
unusual temperature and bias
dependence of current. The observed switching in many cases is \emph{%
extrinsic}, caused by changes in molecule-electrode geometry, molecule
reconfiguration, metallic filament formation through, and/or changing amount
of disorder in a molecular film. We give experimental examples of telegraph
"switching" and "hot spot" formation in the molecular films.

\section{Introduction}

\label{Int}


Current interest in molecular electronics is largely driven by expectations
that molecules can be used as nanoelectronics components able to
complement/replace standard silicon CMOS\ technology\cite{tour00,joachim02}
on the way down to $\leq $10nm circuit components. The first speculations
about molecular electronic devices (diodes, rectifiers) were apparently made
in mid-1970s\cite{Aviram}. That original suggestion of a molecular rectifier
has generated a large interest in the field and a flurry of suggestions of
various molecular electronics components, especially coupled with premature
estimates that silicon-based technology cannot scale to below 1$\mu $m
feature size. The Aviram-Ratner's Donor-insulator-Acceptor construct TTF$%
-\sigma -$TCNQ ($D^{+}-\sigma -A^{-},$ see details below), where carriers
were supposed to tunnel asymmetrically in two directions through insulating
saturated molecular $\sigma -`$bridge', has never materialized, in spite of
extensive experimental effort over a few decades \cite{Martin}. End result
in some cases appears to be a slightly electrically anisotropic \emph{%
insulator,} rather than a diode, unsuitable as a replacement for silicon
devices. This comes about because in order to assemble a reasonable quality
monolayer of these molecules in Langmuir-Blodgett trough (avoiding defects
that will short the device after electrode deposition) one needs to attach a
long `tail' molecule C18 [$\equiv ($CH$_{2})_{18}]$ that can produce enough
of a Van-der-Waals force to keep molecules together, but C18 is a wide-band
insulator with a bandgap $E_{g}\approx 9-10$eV. The outcome of these studies
may have been anticipated, but if one were able to assemble the
Aviram-Ratner molecules without the tail, they could not rectify anyway.
Indeed, a recent ab-initio study\cite{noaviram} of $D^{+}\sigma A^{-}$
prospective molecule showed no appreciable asymmetry of its I-V curve. The
molecule was envisaged by Ellenbogen and Love\cite{ellen00} as a 4-phenyl
ring Tour wire with dimethylene insulating bridge in the middle directly
connected to Au electrodes via thiol groups. Donor-acceptor asymmetry was
produced by side NH$_{2}^{+}$ and NO$_{2}^{-}$ moieties, which is a frequent
motif in molecular devices using the Tour wires. The reason of poor
rectification is simple: the bridge is too short, it is a transparent piece
of one-dimensional insulator, whereas the applied field is three dimensional
and it cannot be screened efficiently with an appreciable voltage drop on
the insulating group in this geometry. Although there is only 0.7eV energy
separation between levels on the D and A groups, one needs about 4eV\ bias
to align them and get a relatively small current because total resonant
transparency is practically impossible to achieve. Remember, that the model
calculation implied an ideal coupling to electrodes, which is impossible in
reality and which is known to dramatically change the current through the
molecule (see below). We shall discuss below some possible alternatives to
this approach.

It is worth noting that studies of energy and electron transport in
molecular crystals \cite{davydov62} started already in early 1960s. It was
established in mid-1960s in what circumstances charge transport in
biological molecules involves electron tunneling \cite{gutmann68}. It was
realized in mid-1970s that since the organic molecules are `soft', energy
transport along linear biological molecules, proteins, etc. may proceed by
low energy nonlinear collective excitations, like Davydov solitons \cite
{davydov77} (see review \cite{DavSolRev92}).

To take over from current silicon CMOS technology, the molecular
electronics should provide smaller, more reliable, functional
components that can be produced and assembled concurrently and are
compatible with CMOS for integration. The small size of units that
molecules may hopefully provide is quite obvious. However, meeting
other requirements seems to be a very long shot. To beat alternative
technologies for e.g. dense (and cheap) memories, one should aim at
a few Tb/in$^{2}$ ($>10^{12}-10^{13}$ bit/cm$^{2})$, which
corresponds to linear bit (footprint)\ sizes of $3-10$ nanometers,
and an operation lifetime of $\sim 10$ years. The latter requirement
is very difficult to meet with organic molecules that tend to
oxidize and decompose, especially under conditions of very high
applied electric field (given the operational voltage bias of $\sim
1$V for molecules integrated with CMOS and their small sizes on the
order of a few nanometers). In terms of areal density, one should
compare this with rapidly developing technologies like ferroelectric
random access (FERAM)\cite{scott00} or phase-change memories
(PCM)\cite{intel06}. The current smallest commercial
nano-ferroelectrics are
about 400$\times $400 nm$^{2}$ and 20-150 nm thick \cite{jung05}, and the $%
128\times 128$ arrays of switching ferroelectric pixels bits have been
already demonstrated with a bit size $\leq 50$nm (with density $\sim $Tb/in$%
^{2})$\cite{50nmFE}. The phase-change memories based on
chalcolgenides GeSbTe (GST) seem to scale even better than the
ferroelectrics. As we see, the mainstream technology for
random-access memory approaches molecular size very rapidly. At the
same time, the so-called ``nanopore'' molecular devices\cite
{ReedRAM} have comparable sizes and yet to demonstrate a repeatable
behavior (for reasons explained below).

In terms of parallel fabrication of molecular devices, one is looking at
\emph{self-assembly} techniques (see, e.g. \cite{ted,brus03}, and references
therein). Frequently, the Langmuir-Blodgett technique is used for
self-assembly of molecules on water, where molecules are prepared to have
hydrophilic ``head'' and hydrophobic ``tail'' to make the assembly possible,
see e.g. Refs.~\cite{petty,Ulman95}. The allowances for a corresponding
assembly, especially of hybrid structures (molecules integrated on silicon
CMOS), are on the order of a \emph{fraction} of an Angstrom, so actually a
\emph{picotechnology }is required \cite{joachim02}. Since it is problematic
to reach such a precision any time soon, the all-in-one molecule approach
was advocated, meaning that a fully functional computing unit should be
synthesized as a single supermolecular unit\cite{joachim02}. The hope is
that perhaps directed self-assembly will help to accomplish building such a
unit, but self-assembly on a large scale is impossible without defects\cite
{ted,brus03}, since the entropic factors work against it. Above some small
defect concentration (``percolation'') threshold the mapping of even a
simple algorithm on such a self-assembled network becomes impossible\cite
{greg04}.

There is also a big question about electron transport in such a
device consisting of large organic molecules. Even in high-quality
pentacene (P5) crystals, perhaps the best materials for thin film
transistors, the mobility is a mere 1-2 cm$^{2}$/V$\cdot $s (see
e.g. \cite{pentacene05}), as a result of carrier trapping by
interaction with a lattice and necessity to hop between P5
molecules. The situation with carrier transport through long
molecules ($>2-3$nm) is, of course, substantially different from the
transport through short rigid molecules that have been envisaged as
possible electronics components. Indeed, in \emph{short} molecules
the dominant mode of electron transport would be resonant tunneling
through \emph{electrically active} molecular orbital(s)
\cite{restunn}, which, depending on the workfunction of the
electrode, affinity of the molecule, and symmetry of coupling
between molecule and electrode may be one of the lowest unoccupied
molecular orbitals (LUMO)\ or highest occupied molecular orbitals
(HOMO)\cite{BK03,KB01}. Indeed, it is well known that in longer
wires containing more than about 30-40 atomic sites, the tunneling
time is comparable to or larger than the characteristic phonon
times, so that the polaron (and/or bipolaron) can be formed inside
the molecular wire \cite{fis}. There is a wide range of molecular
bulk conductors with (bi)polaronic carriers. The formation of
polarons (and charged solitons) in polyacetylene (PA)\ was discussed
a long time ago theoretically in Refs.~\cite{su80} and formation of
bipolarons (bound states of two polarons) in Ref.\cite{braz81}.
Polarons in PA were detected optically in Ref.\cite{feldblum82} and
since then studied in great detail. There is an exceeding amount of
evidence of the polaron and bipolaron formation in conjugated
polymers such as polyphenylene, polypyrrole, polythiophene,
polyphenylene sulfide \cite{bredas84}, Cs-doped biphenyl \cite
{ramsey90}, n-doped bithiophene \cite{steinmuller93},
polyphenylenevinylene(PPV)-based light emitting diodes
\cite{swanson93}, and other molecular systems. Both intrinsic and
extrinsic parameters play a role in determining the electrical and
optical properties of polymer films: spatial range of $\pi$-electron
delocalization, interchain interaction, morphology, amount of
defects and disorder, carrier density, etc (for a brief review of
carrier transport mechanisms and materials see
Ref.~\cite{jaiswal06}).

The latest wave of interest in molecular electronics is mostly related to
recent studies of carrier transport in synthesized linear conjugated
molecular wires (Tour wires\cite{tour00}) with apparent non-linear I-V\
characteristics [negative differential resistance (NDR)] and ``memory''
effects\cite{ReedNDR,donhauser,ReedRAM}, various molecules with a mobile
\emph{microcycle} that is able to move back and forth between metastable
conformations in solution (molecular shuttles) \cite{rotax} and demonstrate
some sort of \ ``switching'' between relatively stable resistive states when
sandwiched between electrodes in a solid state device\cite{yong03} (see also
\cite{DSindSw04}). There are also various photochromic molecules that may
change conformation (``switch'')\ upon absorption of light \cite
{molswitchbook01}, which may be of interest to some photonics applications
but not for the general purpose electronics. One of the most serious
problems with using this kind of molecules is \emph{power dissipation}.
Indeed, the studied organic molecules are, as a rule, very resistive (in the
range of $\sim 1$M$\Omega -1$G$\Omega $, or more). Since usually the
switching bias exceeds $0.5$V the dissipated power density would be in
excess of 10 kW/cm$^{2},$ which is orders of magnitude higher than the
presently manageable level. One can drop the density of switching devices,
but this would undermine a main advantage of using molecular size elements.
This is a common problem that CMOS\ faces too, but organic molecules do not
seem to offer a tangible advantage yet. There are other outstanding
problems, like understanding an actual switching mechanism, which seems to
be rather molecule-independent\cite{DSindSw04}, stability, scaling, etc. It
is not likely, therefore, that molecules will displace silicon technology,
or become a large part of a hybrid technology in a foreseeable future.

First major moletronic applications would most likely come in the area of
chemical and biological sensors. One of the current solutions in this area
is to use the functionalized nanowires. When a target analyte molecule
attaches from the environment to such a nanowire, it changes the
electrostatic potential ``seen''\ by the carriers in the nanowire. Since the
conductance of the nanowire device is small, even one chemisorbed molecule
could make a detectable change of a conductance\cite{lieberNWS}.
Semiconducting nanowires can be grown from seed metal nanoparticles\cite{ted}%
, or it can be carbon nanotubes (CNT), which are studied extensively due to
their relatively simple structure and some unique properties like very high
conductance\cite{CNT}.

In this paper we shall address various generic problems related to electron
transport through molecular devices, and describe some specific molecular
systems that may be interesting for applications as rectifiers and switches,
and some pertaining physical problems. We shall first consider systems where
an elastic tunneling is dominant, and interaction with vibrational
excitations on the molecules only renormalizes some parameters describing
tunneling. We shall also describe a situation where the coupling of carriers
to molecular vibrons is strong. In this case the tunneling is substantially
inelastic and, moreover, it may result in current hysteresis when the
electron-vibron interaction is so strong that it overcomes Coulomb repulsion
of carriers on a central narrow-band/conjugated unit of the molecule
separated from electrodes by wide band gap saturated molecular groups like
(CH$_{2}$)$_{n},$ which we shall call a molecular quantum dot (molQD).
Another very important problem is to understand the nature and the role of
imperfections in organic thin films. It is addressed in the last section of
the paper.

\section{Role of molecule-electrode contact: extrinsic molecular switching
due to molecule tilting}

We have predicted some while ago that there should be a strong dependence of
the current through conjugated molecules (like the Tour wires \cite{tour00})
on the geometry of molecule-electrode contact \cite{BK03,KB01}. The apparent
``telegraph'' switching observed in STM\ single-molecule probes of the
three-ring Tour molecules, inserted into a SAM of non-conducting shorter
alkanes, has been attributed to this effect \cite{donhauser}. The theory
predicts very strong dependence of the current through the molecule on the
tilting angle between a backbone of a molecule and a normal to the electrode
surface. Other explanations, like rotation of the middle ring, charging of
the molecule, or effects of the moieties on the middle ring, do not hold. In
particular, switching of the molecules \emph{without} any NO$_{2}$ or NH$_{2}
$ moieties have been practically the same as with them.

The simple argument in favor of the ``tilting'' mechanism of the conductance
lies in a large anisotropy of the molecule-electrode coupling through $\pi $%
-conjugated molecular orbitals (MOs). In general, we expect the overlap and
the full conductance to be maximal when the lobes of the $p$-orbital of the
end atom at the molecule are oriented perpendicular to the surface, and
smaller otherwise, as dictated by the symmetry. The overlap integrals of a $%
p $-orbital with orbitals of other types differ by a factor about 3 to 4 for
the two orientations. Since the conductance is proportional to the square of
the matrix element, which contains a product of two metal-molecule hopping
integrals, the total conductance variation with overall geometry may
therefore reach two orders of magnitude, and in special cases be even larger.

\begin{figure}[ptb]
\begin{center}
\includegraphics [angle=0, width=.75\textwidth]
{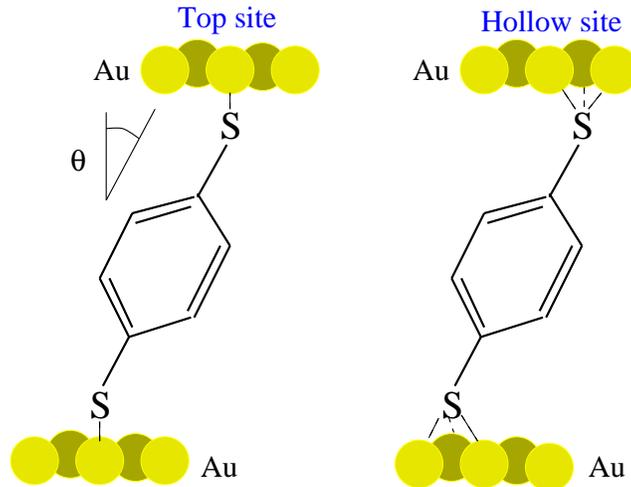}
\end{center}
\caption{Schematic representation of the benzene-dithiolate molecule on top
and hollow sites. End sulfur atoms are bonded to one and three surface gold
atoms, respectively, $\protect\theta $ is the tilting angle.}
\label{fig:bdt}
\end{figure}

In order to illustrate the geometric effect on current we have considered a
simple two-site model with $p-$orbitals on both sites, coupled to electrodes
with $s-$orbitals \cite{KB01}. For non-zero bias the transmission
probability has the resonant form (\ref{eq:TEbw}) with line widths for
hopping to the left (right) lead $\Gamma _{L}$ $\left( \Gamma _{R}\right) .$
The current has the approximate form (with $\Gamma =\Gamma _{L}+\Gamma _{R})$
\begin{equation}
I\approx \left\{
\begin{array}{cc}
\frac{q^{2}}{h}\frac{\Gamma ^{2}}{t_{\pi }^{2}}V\propto \sin ^{4}\theta , &
qV\ll E_{\mathrm{LUMO}}-E_{\mathrm{HOMO}}, \\
\frac{8\pi q}{h}\frac{\Gamma _{L}\Gamma _{R}}{\Gamma _{L}+\Gamma _{R}}%
\propto \sin ^{2}\theta , & qV>E_{\mathrm{LUMO}}-E_{\mathrm{HOMO}},
\end{array}
\right.  \label{eq:I2case}
\end{equation}
where $\theta $ is the tilting angle\cite{KB01,BK03}, Fig.~\ref{fig:bdt}.

\begin{figure}[h]
\begin{center}
\includegraphics [angle=0, width=.75\textwidth]
{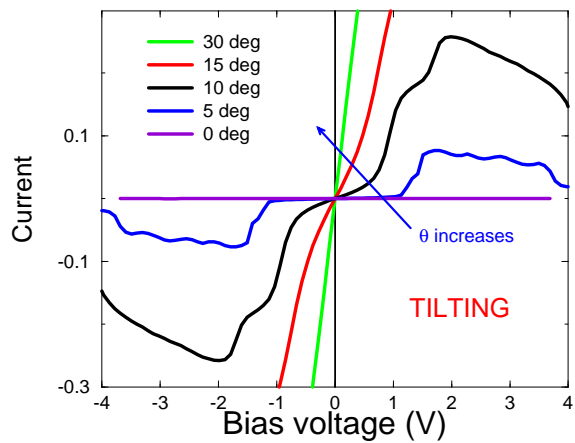}
\end{center}
\caption{ Effect of tilting on I-V curve of the BDT molecule, Fig.~\ref
{fig:bdt}. Current is in units of $I_{0}=77.5\protect\mu \mathrm{A}$, $%
\protect\theta $ is the tilting angle. }
\label{fig:IVbdt}
\end{figure}

The tilting angle has a large effect on the I-V curves of benzene-dithiolate
(BDT) molecules, especially when the molecule is anchored to the Au
electrode in the top position, Fig.~\ref{fig:IVbdt}. By changing $\theta $
from 5$^{\circ }$ to just 15$^{\circ }$, one drives the I-V characteristic
from the one with a gap of about 2V to the ohmic one with a large relative
change of conductance. Even changing $\theta $ from 10$^{\circ }$ to 15$%
^{\circ }$ changes the conductance by about an order of magnitude. The I-V
curve for the hollow site remains ohmic for tilting angles up to 75$^{\circ
} $ with moderate changes of conductance. Therefore, if the molecule in
measurements snaps from the top to the hollow position and back, it will
lead to an apparent switching \cite{donhauser}. It has recently been
realized that the geometry of a contact strongly affects coherent spin
transfer between molecularly bridged quantum dots \cite{molmolspin}. It is
worth noting that another frequently observed \emph{extrinsic} mechanism of
``switching''\ in organic layers is due to electrode material diffusing into
the layer and forming metallic ``filaments'' (see below).

\section{Molecular quantum dot rectifiers}

Aviram and Ratner speculated about a rectifying molecule containing donor ($%
D)$ and acceptor ($A)$ groups separated by a saturated $\sigma $-bridge
(insulator) group, where the (inelastic) electron transfer will be more
favorable from $A$ to $D$ \cite{Aviram}. The molecular rectifiers actually
synthesized, C$_{16}$H$_{33}-\gamma $Q3CNQ, were of somewhat different $%
D-\pi -A$ type, i.e. the ``bridge'' group was conjugated \cite{Martin}.
Although the molecule did show rectification (with considerable hysteresis),
it performed rather like an anisotropic insulator with tiny currents on the
order of $10^{-17}$A/molecule, because of the large alkane ``tail'' needed
for LB\ assembly. It was recently realized that in this molecule the
resonance does not come from the alignment of the HOMO and LUMO, since they
cannot be decoupled through the conjugated $\pi -$bridge, but rather due to
an asymmetric voltage drop across the molecule \cite{Krzeminski}. Rectifying
behavior in other classes of molecules is likely due to asymmetric contact
with the electrodes \cite{Zhou,xue99}, or an asymmetry of the molecule
itself \cite{reichert}. To make rectifiers, one should avoid using molecules
with long insulating groups, and we have suggested using relatively short
molecules with ''anchor'' end groups for their self-assembly on a metallic
electrodes, with a phenyl ring as a central conjugated part \cite{KBWmr02}.
This idea has been tested in Ref.~\cite{krze2003} with a phenyl and
thiophene rings attached to a (CH$_{2}$)$_{15}$ tail by a CO group. The
observed rectification ratio was $\leq $10, with some samples showing the
ratio of about 37.

\begin{figure}[h]
\begin{center}
\includegraphics [angle=0, width=1\textwidth]
{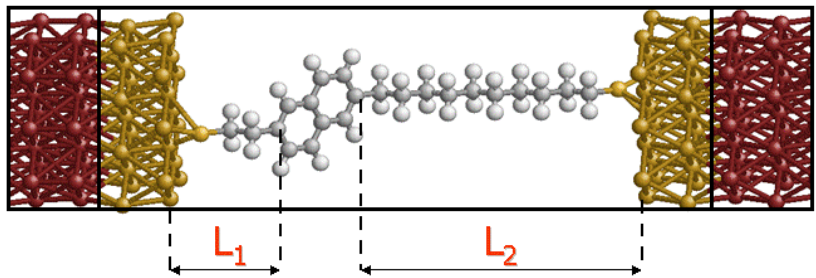}
\end{center}
\caption{ Stick figure representing the naphthalene conjugated central unit
separated from the left(right) electrode by saturated (wide band gap) alkane
groups with length $L_{1(2)}$. }
\label{fig:naph}
\end{figure}

We have recently studied a more promising rectifier like $-$S-(CH$_{2})_{2}$%
-Naph$-\newline
$(CH$_{2})_{10}-$S$-$ with a theoretical rectification $\leq 100$\cite
{larade03}, Fig.~\ref{fig:naph}. This system has been synthesized and
studied experimentally \cite{shunchi}. To obtain an accurate description of
transport in this case, we employ an ab-initio non-equilibrium Green's
function method \cite{Guo01}. The present calculation takes into account
only elastic tunneling processes. Inelastic processes may substantially
modify the results in the case of strong interaction of the electrons with
molecular vibrations, see Ref.~\cite{AB03vibr} and below. There are
indications in the literature that the carrier might be trapped in a polaron
state in saturated molecules somewhat longer than those we consider in the
present paper \cite{boulas96}. One of the barriers in the present rectifiers
is short and relatively transparent, so there will be no appreciable Coulomb
blockade effects. The structure of the present molecular rectifier is shown
in Fig.~\ref{fig:naph}. The molecule consists of a central conjugated part
(naphthalene) isolated from the electrodes by two insulating aliphatic
chains (CH$_{2}$)$_{n}$ with lengths $L_{1}$($L_{2}$) for the left (right)
chain.

\begin{figure}[h]
\begin{center}
\includegraphics [angle=0, width=.75\textwidth]
{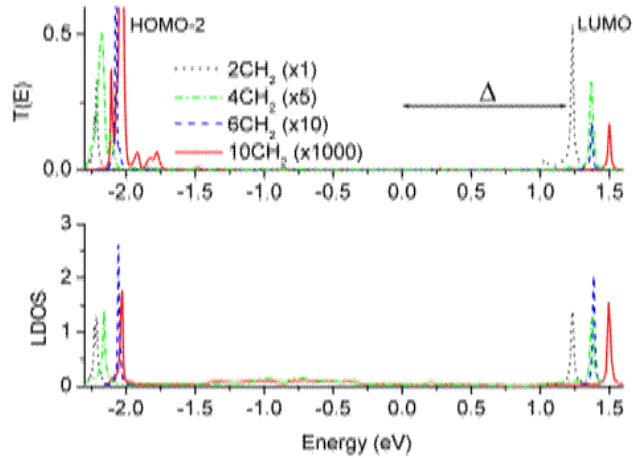}
\end{center}
\caption{ Transmission coefficient versus energy E for rectifiers $-$S-(CH$%
_{2})_{2}$-C$_{10}$H$_{6}-$(CH$_{2})_{n}-$S$-,$ $n=2,4,6,10$. $\Delta$
indicates the distance of the closest MO to the electrode Fermi energy (E$%
_{F}$ = 0). }
\label{fig:Trannaph}
\end{figure}

\begin{figure}[h]
\begin{center}
\includegraphics [angle=0, width=.75\textwidth]
{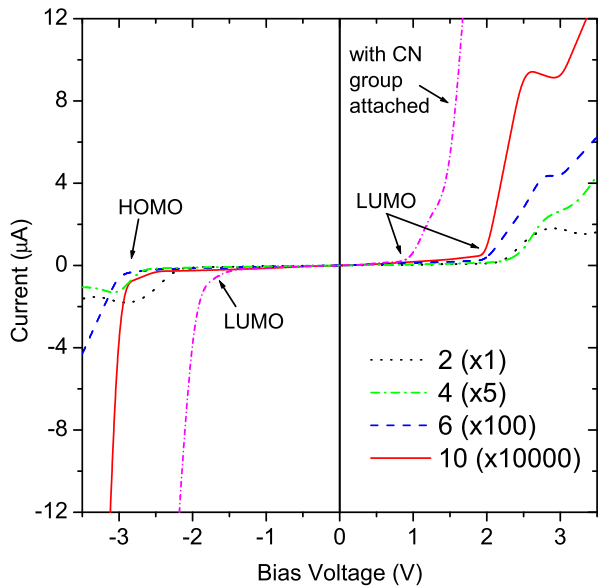}
\end{center}
\caption{I-V curves for naphthalene rectifiers $-$S-(CH$_{2})_{2}$-C$_{10}$H$%
_{6}-$(CH$_{2})_{n}-$S$-,$ $n=2,4,6,10$. The short-dash-dot curve
corresponds to a cyano-doped (added group -C$\equiv$N) $n=10$ rectifier. }
\label{fig:IVnaph}
\end{figure}

The principle of molecular rectification by a molecular quantum dot is
illustrated in Fig.~\ref{fig:Trannaph}, where the electrically ``active''
molecular orbital, localized on the middle conjugated part, is the LUMO,
which lies at an energy $\Delta$ above the electrode Fermi level at zero
bias. The position of the LUMO is determined by the work function of the
metal $q\phi$ and the affinity of the molecule $q\chi,$ $\Delta=\Delta_{%
\mathrm{LUMO}}=q\left( \phi -\chi\right) .$ The position of the HOMO\ is
given by $\Delta_{\mathrm{HOMO}}=\Delta_{\mathrm{LUMO}}-E_{g},$ where $E_{g}$
is the HOMO-LUMO\ gap. If this orbital is considerably closer to the
electrode Fermi level $E_{F},$ then it will be brought into resonance with $%
E_{F}$ prior to other orbitals. It is easy to estimate the forward and
reverse bias voltages, assuming that the voltage mainly drops on the
insulating parts of the molecule,
\begin{eqnarray}
& V_{F} =\frac{\Delta}{q}(1+\xi),\qquad V_{R}=\frac{\Delta}{q}\left( 1+\frac{%
1}{\xi}\right) , \\
& V_{F}/V_{R} =\xi\equiv L_{1}/L_{2},  \label{eq:vratio}
\end{eqnarray}
where $q$ is the elementary charge. A significant difference between forward
and reverse currents should be observed in the voltage range $%
V_{F}<|V|<V_{R} $. The current is obtained from the Landauer formula
\begin{equation}
I=\frac{2q^{2}}{h}\int dE\left[ f(E)-f(E+qV)\right] g(E,V).  \label{eq:cur}
\end{equation}
We can make qualitative estimates in the resonant tunneling model, with the
conductance $g(E,V)\equiv T(E,V)/q,$ where $T(E,V)\;$is the transmission
given by the Breit-Wigner formula
\begin{equation}
T(E,V)=\frac{\Gamma _{L}\Gamma _{R}}{\left( E-E_{MO}\right) ^{2}+\left(
\Gamma _{L}+\Gamma _{R}\right) ^{2}/4},  \label{eq:TEbw}
\end{equation}
$E_{MO}$ is the energy of the molecular orbital. The width $\Gamma
_{L(R)}\sim t^{2}/D=\Gamma _{0}e^{-2\kappa L_{1(2)}},$ where $t$ is the
overlap integral between the MO\ and the electrode, $D$ is the electron band
width in the electrodes, $\kappa $ the inverse decay length of the resonant
MO into the barrier. The current above the resonant threshold is
\begin{equation}
I\approx \frac{2q}{\hbar }\Gamma _{0}e^{-2\kappa L_{2}}.
\end{equation}
We see that increasing the spatial asymmetry of the molecule ($L_{2}/L_{1})$
changes the operating voltage range linearly, but it also brings about an
\emph{exponential} decrease in current \cite{KBWmr02}. This severely limits
the ability to optimize the rectification ratio while simultaneously keeping
the resistance at a reasonable value.
To calculate the I-V curves, we use an ab-initio approach that combines the
Keldysh non-equilibrium Green's function (NEGF) with pseudopotential-based
real space density functional theory (DFT) \cite{Guo01}. The main advantages
of our approach are (i) a proper treatment of the open boundary condition;
(ii) a fully atomistic treatment of the electrodes and (iii) a
self-consistent calculation of the non-equilibrium charge density using
NEGF. The transport Green's function is found from the Dyson equation
\begin{equation}
\left( G^{R}\right) ^{-1}=\left( G_{0}^{R}\right) ^{-1}-V,
\end{equation}
where the unperturbed retarded Green's function is defined in operator form
as $\left( G_{0}^{R}\right) ^{-1}=(E+i0)\hat{S}-\hat{H},$ $H$ is the
Hamiltonian matrix for the scatterer (molecule plus screening part of the
electrodes). $S$ is the \emph{overlap} matrix, $S_{i,j}=\left\langle \chi
_{i}|\chi _{j}\right\rangle $ for non-orthogonal basis set orbitals $\chi
_{i}$, and the coupling of the scatterer to the leads is given by the
Hamiltonian matrix $V=\mathrm{diag}[\Sigma _{l,l},~0,~\Sigma _{r,r}],$ where
$l$ ($r$) stands for left (right) electrode. The self-energy part $\Sigma
^{<},$ which is used to construct the non-equilibrium electron density in
the scattering region, is found from $\Sigma ^{<}=-2i\mathrm{Im}\left[
f(E)\Sigma _{l,l}+f(E+qV)\Sigma _{r,r}\right] ,$ where $\Sigma _{l,l(r,r)}$
is the self-energy of the left (right) electrode, calculated for the
semi-infinite leads using an iterative technique \cite{Guo01}. \ $\Sigma ^{<}
$ accounts for the steady charge ``flowing in'' from the electrodes. The
transmission probability is given by
\begin{equation}
T(E,V)=4\mathrm{Tr}\left[ \left( \rm{Im}\Sigma _{l,l}\right)
G_{l,r}^{R}\left( \rm{Im}\Sigma _{r,r}\right) G_{r,l}^{A}\right] ,
\label{eq:gnegf}
\end{equation}
where $G^{R(A)}$ are the retarded (advanced)\ Green's function, and $\Sigma $
the self-energy part connecting left ($l$) and right ($r$) electrodes \cite
{Guo01}, and the current is obtained from Eq.~(\ref{eq:cur}). The calculated
transmission coefficient $T(E)$ is shown for a series of rectifiers $-$S-(CH$%
_{2})_{m}$-C$_{10}$H$_{6}-$(CH$_{2})_{n}-$S$-$ for $m=2$ and $n=2,4,6,10$ at
zero bias voltage in Fig.~\ref{fig:Trannaph}. We see that the LUMO is the
molecular orbital transparent to electron transport, lies above $E_{F}$ by
an amount $\Delta=1.2-1.5$eV. The transmission through the HOMO\ and HOMO-1
states, localized on the terminating sulfur atoms, is negligible, but the
HOMO-2 state conducts very well. The HOMO-2 defines the threshold reverse
voltage $V_{R},$ thus limiting the operating voltage range. Our assumption,
that the voltage drop is proportional to the lengths of the alkane groups on
both sides, is quantified by the calculated potential ramp. It is close to a
linear slope along the (CH$_{2})_{n}$ chains \cite{larade03}. The forward
voltage corresponds to the crossing of the LUMO($V)$\ and $\mu _{R}(V)$,
which happens at about 2V. Although the LUMO\ defines the forward threshold
voltages in all molecules studied here, the reverse voltage is defined by
the HOMO-2\ for ``right'' barriers (CH$_{2})_{n}$ with $n=6,10$. The I-V
curves are plotted in Fig.~\ref{fig:IVnaph}.
We see that the rectification ratio for current in the operation window $%
I_{+}/I_{-}$ reaches a maximum value of 35 for the ``2-10'' molecule ($m=2,$
$n=10).$ Series of molecules with a central \emph{single phenyl} ring \cite
{KBWmr02} do not show any significant rectification. One can manipulate the
system in order to increase the energy asymmetry of the conducting orbitals
(reduce $\Delta)$. To shift the LUMO towards $E_{F}$, one can attach an
electron withdrawing group, like $-$C$\equiv$N \cite{larade03}. The
molecular rectification ratio is not great by any means, but one should bear
in mind that this is a device necessarily operating in a ballistic
quantum-mechanical regime because of the small size. This is very different
from present Si devices with carriers diffusing through the system. As
silicon devices become smaller, however, the same effects will eventually
take over, and tend to diminish the rectification ratio, in addition to
effects of finite temperature and disorder in the system.

\section{Molecular switches}

There are various molecular systems that exhibit some kind of current
``switching'' behavior \cite{donhauser,rotax,yong03,DSindSw04}, ``negative
differential resistance''\cite{ReedNDR}, and ``memory''\cite{ReedRAM}. The
switching systems are basically driven between two states with considerably
different resistances. This behavior is not really sensitive to a particular
molecular structure, since this type of bistability is observed in complex
rotaxane-like molecules as well as in very simple alkane chains (CH$_{2}$)$%
_{n}$ assembled into LB\ films \cite{DSindSw04}, and is not even exclusive
to the organic films. The data strongly indicates that the switching has an
extrinsic origin, and is related either to bistability of molecule-electrode
orientation \cite{BK03,KB01,donhauser}, or transport assisted by defects in
the film\cite{polymer78,tredgold81}.

\subsection{Extrinsic switching in organic molecular films: role of defects
and molecular reconfigurations}

Evidently, large defects can be formed in organic thin films as a result of
electromigration in very strong field, as was observed long ago\cite
{LBfila8671}. It was concluded some decades ago that the conduction through
absorbed \cite{polymer78} and Langmuir-Blodgett \cite{tredgold81} monolayers
of fatty acids (CH$_{2})_{n},$ which we denote as Cn, is associated with
\emph{defects}. In particular, Polymeropoulos and Sagiv studied a variety of
absorbed monolayers from C7 to C23 on Al/Al$_{2}$O$_{3}$ substrates and
found that the exponential dependence on the length of the molecular chains
is only observed below the liquid nitrogen temperature of 77K, and no
discernible length dependence was observed at higher temperatures \cite
{polymer78}. The temperature dependence of current was strong, and was
attributed to transport assisted by some defects. The current also varied
strongly with the temperature in Ref.~\cite{tredgold81} for LB films on Al/Al%
$_{2}$O$_{3}$ substrates in He atmosphere, which is not compatible with
elastic tunneling. Since the He atmosphere was believed to hinder the Al$%
_{2} $O$_{3}$ growth, and yet the resistance of the films increased about
100-fold over 45 days, the conclusion was made that the ``defects'' somehow
anneal out with time. Two types of switching have been observed in 3-30 $\mu
$m thick films of polydimethylsiloxane (PDMS), one as a standard dielectric
breakdown with electrode material ``jet evaporation'' into the film with
subsequent Joule melting of metallic filament under bias of about 100V, and
a low-voltage ($<$1 V) ``ultraswitching'' that has a clear ``telegraph''
character and resulted in intermittent switching into a much more conductive
state\cite{shlimak98}. The exact nature of this switching also remains
unclear, but there is a strong expectations that the formation of metallic
filaments that may even be in a ballistic regime of transport, may be
relevant to the phenomenon.

\begin{figure}[t]
\begin{center}
\includegraphics [angle=0, width=.75\textwidth]
{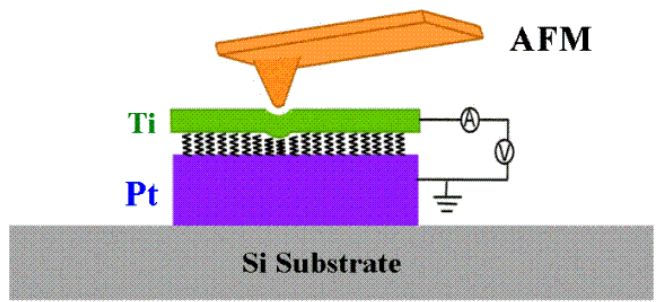}
\end{center}
\caption{ Experimental setup for mapping local conductance. AFM produces
local deformation of top electrode and underlying organic film. The \emph{%
total} conductance of the device is measured and mapped. (Courtesy C.N.
Lau). }
\label{fig:j1}
\end{figure}

Recently, a direct evidence was obtained of the formation of ``hot spots''
in the LB\ films that may be related to the filament growth through the film
imaged with the use of AFM\ current mapping\cite{Jennie04}. The system
investigated in this work has been Pt/stearic acid (C18)/Ti (Pt/C18/Ti)
crossbar molecular structure, consisting of planar Pt and Ti electrodes
sandwiching a monolayer of 2.6-nm-long stearic acid (C18H36OH) molecules
with typical zero-bias resistance in excess of $10^{5}\Omega $. The devices
has been switched reversibly and repeatedly to higher (``on'') or lower
(``off'') conductance states by applying sufficiently large bias voltage $%
V_{b}$ to top Ti electrode with regards to Pt counterelectrode, Fig.~\ref
{fig:j1}.

\begin{figure}[t]
\begin{center}
\includegraphics [angle=0, width=1\textwidth]
{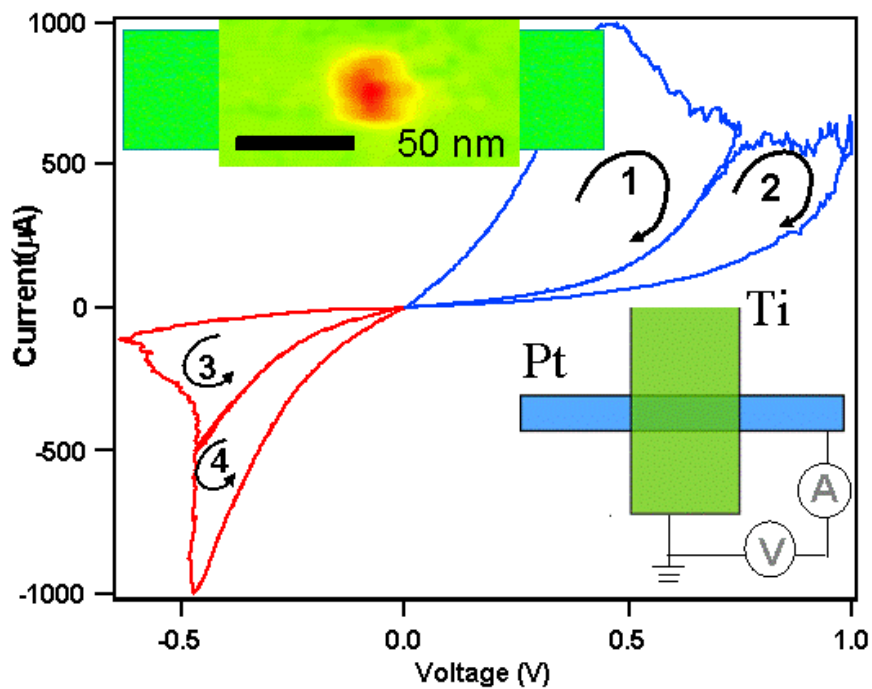}
\end{center}
\caption{ I-V characteristic showing the reversible switching cycle of the
device (bottom inset) with organic film. The arrows indicate sweep
direction. A negative bias switches the device to a high conductance state,
while the positive one switches it to low conductance state. The mapping
according the schematic in Fig.~\ref{fig:j1} shows the appearance of ``hot
spots'' after switching (top inset). (Courtesy C.N. Lau).}
\label{fig:j2}
\end{figure}

Interestingly, reversible switching was not observed in symmetric Pt/C18/Pt
devices. The local conductance maps of the Pt/C18/Ti structure have been
constructed by using an AFM tip and simultaneously measuring the current
through the molecular junction biased to $V_{b}=0.1$V (AFM\ tip was not used
as an electrode, only to apply local pressure at the surface). The study
revealed that the film showed pronounced switching between electrically very
distinct states, with zero-bias conductances $0.17\mu $S (``off'' state) and
1.45$\mu $S (``on'' state), Fig.~\ref{fig:j2}.

At every switching ``on'' there appeared a local conductance peak on the map
with a typical diameter $\sim 40$nm, which then disappeared upon switching
``off'', Fig.~\ref{fig:j2} (top inset). The switching has been attributed to
local conducting \ filament formation due to electromigration processes. It
remains unclear how exactly the filaments dissolve under opposite bias
voltage, why they tend to appear in new places after each switching, and why
conductance in some cases strongly depends on temperature. It is clear,
however, that switching in such a simple molecule without any redox centers,
mobile groups, or charge reception centers should be \emph{extrinsic}.
Interestingly, very similar ``switching'' between two resistive states has
also been observed for tunneling through thin \emph{inorganic} perovskite
oxide films\cite{waser03}.

There have been plenty of reports on non-linear I-V characteristics like
negative differential resistance (NDR) and random switching recently for
molecules assembled on metal electrodes (gold) and silicon. Reports on NDR
for molecules with metal contacts (Au, Hg) have been made in \cite
{ReedNDR,xue99,gorman04}. It became very clear though that most of these
observations are related to molecular reconfigurations and bond breaking and
making, rather to any intrinsic mechanism, like redox states, speculated
about in the original Ref.~\cite{ReedNDR}. Thus, the NDR in Tour wires was
related to molecular reconfiguration with respect to metallic electrode\cite
{BK03,donhauser}, \ NDR in ferrocene-tethered alkyl monolayers \cite
{gorman04} was found to be related to oxygen damage at high voltage\cite
{lindsay05}. Structural changes and bond breaking have been found to result
in NDR\ in experiments with STM\ \cite{gaudioso00,hla03,yang03} and mercury
droplet contacts\cite{cahen04}.

Several molecules, like styrene, have been studied on degenerate Si surface
and showed an NDR behavior\cite{guisinger04}. However, those results have
been carefully checked later and it was found that the styrene molecules do
not exhibit NDR, but rather sporadically switch between states with
different current while held at the same bias voltage (the blinking effect)
\cite{wolkow06}.

\begin{figure}[ptb]
\begin{center}
\includegraphics [angle=0, width=1\textwidth]
{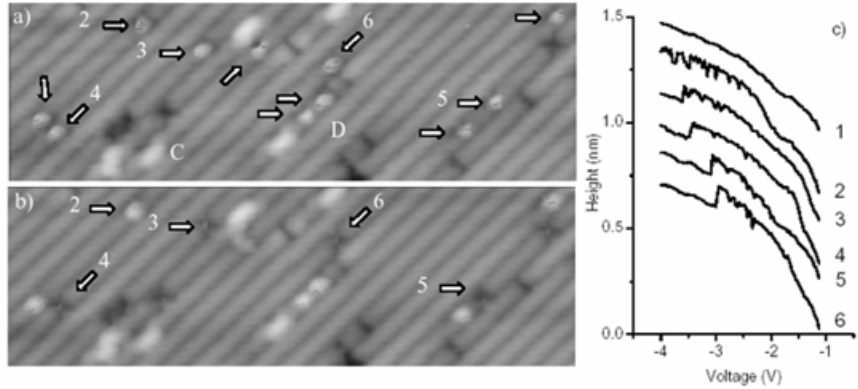}
\end{center}
\caption{STM images or styrene molecules on clean Si(100) before and after
spectroscopy over the area 75 $\times $240 \AA . (Top left) Bias -2~V,
current 0.7~nA. Only the styrene molecules (indicated by arrows) are
blinking during imaging. The clean Si surface, bright defects (marked C),
and dark sports (marked D) do not experience blinking. (Bottom left) bias
-2~V, current 0.3~nA. STM image of individual styrene molecules indicated
with numbers 2-6. Styrene molecules 3-6 have decomposed. Decomposition
involves the changing of the styrene molecule from a bright feature to a
dark depression and also involves the reaction with an adjacent dimer.
Styrene molecule 2 does not decompose and images as usual with no change of
position. (Right) Height-voltage spectra taken over clean silicon (1) and
styrene molecules (2-6). The spectra taken over molecules show several
spikes in height related to blinking in the images. In spectra 3-6, an
abrupt and permanent change in height is recorded and is correlated with
decomposition, as seen in the bottom left image. Spectrum 2 has no permanent
height change, and the molecule does not decompose. (Courtesy J. Pitters and
R. Wolkow). }
\label{fig:w1}
\end{figure}

The STM\ map of the styrene molecules (indicated by arrows) on the Silicon
(100) surface shows that the molecules are blinking, see Fig.~\ref{fig:w1}.
The blinking is absent at clean Si areas, dark (D)\ and bright (C) defects.
This may indicate a dynamic process occurring during the imaging. Comparing
the panel (a) and (b) one may see that some molecules are actually
decomposing. The height versus voltage spectra over particular points are
shown in Fig.~\ref{fig:w1}c. The featureless curve 1 was taken over a clean
silicon dimer. The other spectra were recorded over individual molecules.
Each of these spectra have many sudden decreases and increases in current as
if the molecules are changing between different states during the
measurement causing a change in current and a response of the feedback
control, resulting in a change in height so there exists one or more
configurations that lead to measurement of a different height. Evidently,
these changes have the same origin as the blinking of molecules in STM
images. Figure~\ref{fig:w1}b reveals clear structural changes associated
with those particular spectroscopic changes. In each case where a dramatic
change in spectroscopy occurred, the molecules in the image have changed
from a bright feature to a dark spot. This is interpreted as a decomposition
of the molecules. A detailed look at each decomposed styrene molecule, at
locations 3, 4, 5, and 6, shows that the dark spot is not in precise
registry with the original bright feature, indicating that the decomposition
product involves reaction with an adjacent dimer\cite{wolkow06}.

\begin{figure}[ptb]
\begin{center}
\includegraphics [angle=0, width=1\textwidth]
{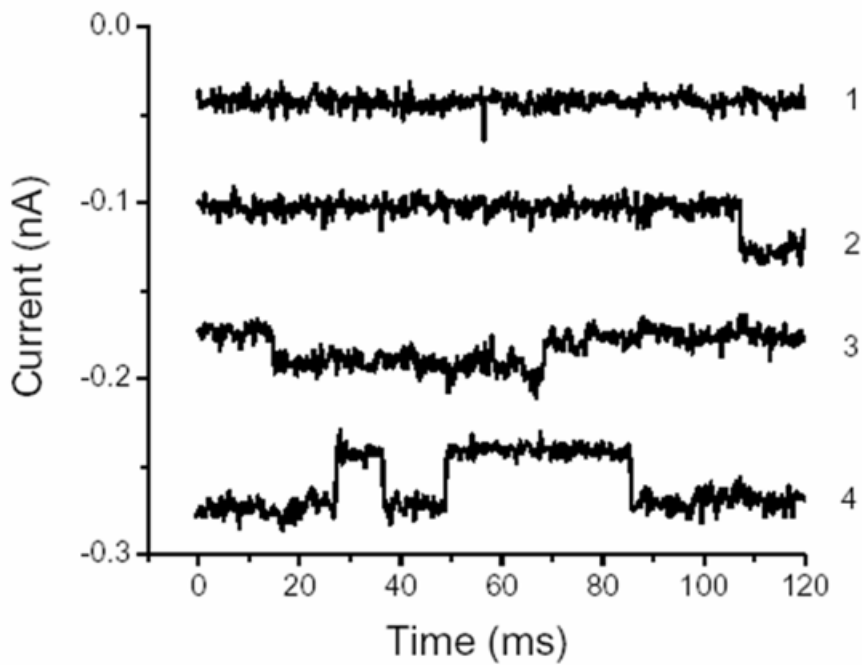}
\end{center}
\caption{Variation of current through styrene molecules on Si(100) with
time. Tunneling conditions were set at -3~V and 0.05~nA. Abrupt increases
and decreases in current relate to changes of the molecule during the
spectroscopy. Some experiments show no changes in current (curve 1), others
show various kinds of telegraph switching. (Courtesy J. Pitters and R.
Wolkow). }
\label{fig:w2}
\end{figure}

The fact that the structural changes and related NDR behavior are not
associated with any resonant tunneling through the molecular levels or redox
processes, but are perhaps related to inelastic electron scattering or other
extrinsic processes, becomes evident from current versus time records shown
in Fig.~\ref{fig:w2}, Ref.\cite{wolkow06}. The records show either no change
of the current with time (1), or one or a few random jumps between certain
current states (telegraph noise). The observed changes in current at a fixed
voltage obviously cannot be explained by shifting and aligning of molecular
levels, as was suggested in Ref.~\cite{dattaNDR04}, they must be related to
an adsorbate molecule structural changes with time. Therefore, the
explanation by Datta \emph{et al.} that the resonant level alignment is
responsible for NDR does not apply\cite{dattaNDR04}. As mentioned above,
similar telegraph switching and NDR has been observed in Tour wires\cite
{donhauser} and other molecules. Therefore, the observed negative
differential resistance apparently has similar origin in disparate molecules
adsorbed on different substrates, and has to do with molecular
reconformation/reconfiguration on the surface.

\subsection{Intrinsic polarization and extrinsic conductance switching in
molecular ferroelectric PVDF}

The only well established, to the best of our knowledge, intrinsic molecular
switching (of polarization, not current) under bias voltage was observed in
molecular ferroelectric block co-polymers polyvinylidene\cite{bune98}.
Ferroelectric polymer films have been prepared with the 70\% vinylidene
fluoride copolymer, P(VDF-TrFE 70:30), formed by horizontal LB deposition on
aluminum-coated glass substrates with evaporated aluminum top electrodes.
The polymer chains contain random sequence of (CH$_{2}$)$_{n}$(CF$_{2}$)$_{m}
$ blocks, fluorine site carries a strong negative charge, and in the
ferroelectric phase most of carbon-fluorine bonds point in one direction.
The fluorine groups can be rotated and aligned in very strong electric
field, $\sim 5$MV/cm. As a result, the whole molecular chain orders, and in
this way the macroscopic polarization can be switched between the opposite
states. The switching process is extremely slow, however, and takes 1-10
seconds (!) \cite{duch00,ALcomm01}. This is not surprising, given strong
Coulomb interaction between charged groups and the metal electrodes, pinning
by surface roughness, and steric hindrance to rotation. This behavior should
be suggestive of other switching systems based on one of few monolayers of
molecules, and other nontrivial behavior involved\cite{ALcomm01}.

The switching of current was also observed in films of PVDF 30 monolayers
thick. The conductance of the film was following the observed hysteresis
loop for the polarization, ranging from $\sim 1\times 10^{-9} - 2\times
10^{-6} \Omega^{-1}$\cite{bune95}. The phenomenon of conductance switching
has these important features: (i) It is connected with the bulk polarization
switching; (ii) there is a large $\sim 1000:1$ contrast between the ON and
OFF states; (iii) the ON state is obtained only when the bulk polarization
is switched in the positive direction; (iv) the conductance switching is
much faster than the bulk polarization switching. The conductance switches
ON only after the 6s delay, after the bulk polarization switching is nearly
complete, presumably when the last layer switches into alignment with the
others, while the conductance switches OFF without a noticeable delay after
the application of reverse bias as even one layer reverses (this may create
a barrier to charge transfer). The slow $\sim 2$s time constant for
polarization switching is probably nucleation limited as has been observed
in high-quality bulk films with low nucleation site densities\cite{furuk84}.
The duration of the conductance switching transition $\sim 2$ms may be
limited only by the much faster switching time of individual layers.

The origin of conductance switching by 3 orders of magnitude is not clear.
It may indeed be related to a changing amount of disorder for
tunneling/hopping electrons. It is conceivable that the carriers are
strongly trapped in polaron states inside PVDF and find optimal path for
hopping in the material, which is incompletely switched. This is an
interesting topic that certainly in need of further experimental and
theoretical study.

\section{Molecular quantum dot switching}

\subsection{ Electrically addressable molecules}

For many applications one needs an \emph{intrinsic} molecular ``switch'',
i.e. a bistable voltage-addressable molecular system with very different
resistances in the two states that can be accessed very quickly. There is a
trade-off between the stability of a molecular state and the ability to
switch the molecule between two states with an external perturbation (we
discuss an electric field, switching involving absorbed photons is
impractical at a nanoscale). Indeed, the applied electric field, on the
order of a typical breakdown field $E_{b}\leq 10^{7}$V/cm, is much smaller
than a typical atomic field $\sim 10^{9}$V/cm, characteristic of the energy
barriers. Small barrier would be a subject for sporadic thermal switching,
whereas a larger barrier $\sim 1-2$eV would be impossible to overcome with
the applied field. One may only change the relative energy of the minima by
external field and, therefore, redistribute the molecules statistically
slightly inequivalently between the two states. An intrinsic disadvantage of
the conformational mechanism, involving motion of ionic group, exceeding the
electron mass by many orders of magnitude, is a slow switching speed ($\sim $%
kHz). In case of supramolecular complexes like rotaxanes and catenanes \cite
{rotax} there are two entangled parts which can change mutual positions as a
result of redox reactions (in solution). Thus, for the rotaxane-based memory
devices a slow switching speed of $\sim 10^{-2}$ seconds was reported.

As a possible conformational $E$-field addressable molecular switch, we have
considered a bistable molecule with -CONH$_{2}$ dipole group \cite{KBWsw02}.
The barrier height is $E_{b}=0.18$eV. Interaction with an external electric
field changes the energy of the minima, but estimated switching field is
huge, $\sim 0.5$V/A. At non-zero temperatures, temperature fluctuations
might result in statistical dipole flipping at lower fields. The I-V curve
shows hysteresis in the 3 to 4 Volts window for two possible conformations.
One can estimate the thermal stability of the state as 58 ps at room
temperature, and 33 ms at 77 K.

We explored a possibility for a fast molecular switching where switching is
due to strong correlation effects on the molecule itself, so-called
molecular quantum dot (MQD).
\begin{figure}[ptb]
\begin{center}
\includegraphics [angle=0, width=1\textwidth]
{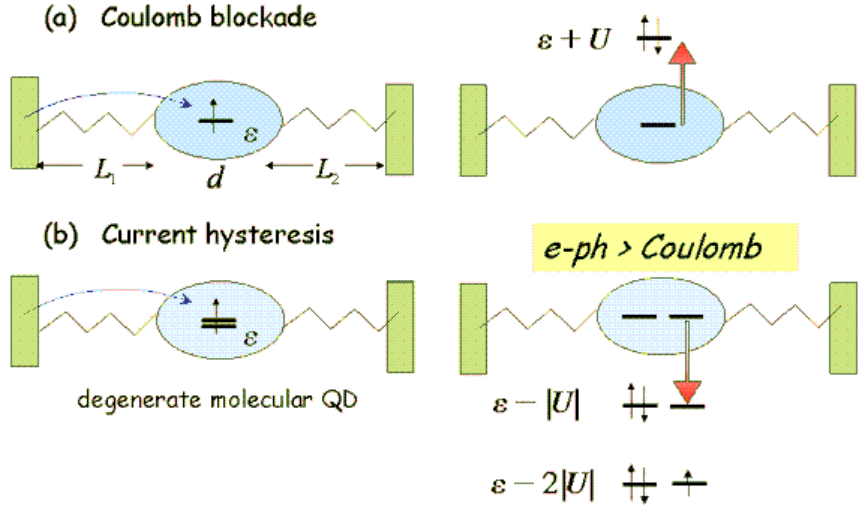}
\end{center}
\caption{Schematic of the molecular quantum dot with central conjugated unit
separated from the electrodes by wide-band insulating molecular groups.
First electron tunnels into the dot and occupies an empty (degenerate) state
there. If the interaction between the first and second incoming electron is
repulsive, $U>0$, then the dot will be in a Coulomb blockade regime (a). If
the electrons on the dot effectively attract each other, $U<0$, the system
will show current hysteresis (b).}
\label{fig:moldot}
\end{figure}
The molecular quantum dot consists of a central \emph{conjugated} unit
(containing half-occupied, and, therefore, extended $\pi -$orbitals), Fig.~%
\ref{fig:moldot}. Frequently, those are formed from the p-states on carbon
atoms, which are not \emph{saturated} (i.e. they do not share electrons with
other atoms forming strong $\sigma -$bonds, with typical bonding-antibonding
energy difference about 1Ry). Since the $\pi -$orbitals are half-occupied,
they form the HOMO-LUMO states. The size of the HOMO-LUMO gap is then
directly related to the size of the conjugated region $d$, Fig.~\ref
{fig:moldot}, by a standard estimate $E_{\text{HOMO-LUMO}}\sim \hbar
^{2}/md^{2}\sim 2-5$~eV. It is worth noting that in the conjugated linear
polymers like polyacetylene ($-\overset{|}{C}=\overset{|}{C}$)$_{n}$ the
spread of the $\pi -$electron would be $d=\infty $ and the expected $E_{%
\text{HOMO-LUMO}}=0.$ However, such a one-dimensional metal is impossible,
Peierls distortion (C=C bond length dimerization) sets in and opens up a gap
of about $\sim 1.5$eV at the Fermi level\cite{petty}. In a molecular quantum
dot the central conjugated part is separated from electrodes by insulating
groups with saturated $\sigma -$bonds, like e.g. the alkane chains, Fig.~3.
Now, there are two main possibilities for carrier transport through the
molQD. If the length of at least one of the insulating groups $L_{1(2)}$ is
not very large (a conductance $G_{1\text{(}2)}$ is not much smaller than the
conductance quantum $G_{0}=2e^{2}/h)$, then the transport through the MQD
will proceed by resonant tunneling processes. If, on the other hand, both
groups are such that the tunnel conductance $G_{1(2)}\ll G_{0},$ the charge
on the dot will be quantized. Then we will have another two possibilities:
(i) the interaction of the extra carriers on the dot is \emph{repulsive} $%
U>0, $ and we have a Coulomb blockade \cite{CoulBlock}, or (ii) the
effective interaction is \emph{attractive}, $U<0,$ then we would obtain the
current \emph{hysteresis} and switching (see below). Coulomb blockade in
molecular quantum dots has been demonstrated in Refs.~\cite{molSET}. In
these works, and in Ref.~\cite{zhitvibr}, the three-terminal active
molecular devices have been fabricated and successfully tested.

Much faster switching compared to the conformational one, described in the
previous section, may be caused by coupling to the vibrational degrees of
freedom, if the vibron-mediated attraction between two carriers on the
molecule is stronger than their direct Coulomb repulsion \cite{AB03vibr},
Fig.~\ref{fig:moldot}b. The attractive energy is the difference of two large
interactions, the Coulomb repulsion and the phonon mediated attraction, on
the order of $1\mathrm{eV}$ each, hence $|U|\sim 0.1$eV.

\subsection{Polaron mechanism of carrier attraction and switching}

Although the correlated electron transport through mesoscopic systems with
repulsive electron-electron interactions received considerable attention in
the past, and continues to be the focus of current studies, much less has
been known about a role of electron-phonon correlations in ``molecular
quantum dots'' (MQD). Some while ago we have proposed a negative$-U$ Hubbard
model of a $d$-fold degenerate quantum dot \cite{alebrawil} and a polaron
model of resonant tunneling through a molecule with degenerate level \cite
{AB03vibr}. We found that the \emph{attractive} electron correlations caused
by any interaction within the molecule could lead to a molecular \emph{%
switching} effect where I-V characteristics have two branches with high and
low current at the same bias voltage. This prediction has been confirmed and
extended further in our theory of \textit{correlated} transport through
degenerate MQDs with a full account of both the Coulomb repulsion and
realistic electron-phonon (e-ph) interactions \cite{AB03vibr}. We have shown
that while the phonon side-bands significantly modify the shape of
hysteretic I-V curves in comparison with the negative-$U$ Hubbard model,
switching remains robust. It shows up when the effective interaction of
polarons is attractive and the state of the dot is multiply degenerate, $d>2$%
.

\begin{figure}[ptb]
\begin{center}
\includegraphics[angle=0, width=.75\textwidth]
{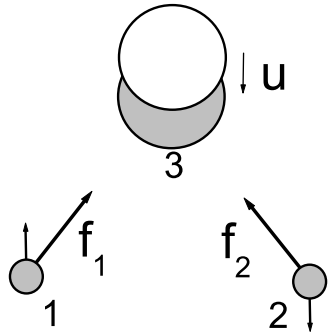} \vskip-0.5mm
\end{center}
\caption{Two localized electrons at sites $\mathbf{1}$ and
$\mathbf{2}$ with opposite spins shift the equilibrium position of
the ion at site $\mathbf{3}$. As a result, the two electrons
\emph{attract} each other.} \label{fig:2sitepol}
\end{figure}

First, we shall describe a simplest model of a single atomic level
coupled with a single one-dimensional oscillator using the first
quantization representation for its displacement $x$ \cite{ABrat06},
\begin{equation}
H=\varepsilon_{0}\hat{n}+fx\hat{n}-{\frac{1}{{2M}}}{\frac{\partial^{2}}{{%
\partial x^{2}}}}+{\frac{kx^{2}}{{2}}}.
\end{equation}
Here $M$ and $k$ are the oscillator mass and the spring constant, $f$ is the
interaction force, and $\hbar=c=k_{B}=1$. This Hamiltonian is readily
diagonalized with the \emph{exact} displacement transformation of the
vibration coordinate $x$,
\begin{equation}
x=y-\hat{n}f/k,
\end{equation}
to the transformed Hamiltonian without electron-phonon coupling,
\begin{align}
\tilde{H} & =\varepsilon\hat{n}-{\frac{1}{{2M}}}{\frac{\partial^{2}}{{%
\partial y^{2}}}}+{\frac{ky^{2}}{{2}}},  \label{eq:newH} \\
\varepsilon & =\varepsilon_{0}-E_{p},   \label{eq:newLev}
\end{align}
where we used $\hat{n}^{2}=\hat{n}$ because of the Fermi-Dirac
statistics. It describes a small polaron at the atomic level
$\varepsilon_{0}$ shifted down by the polaron level shift
$E_{p}=f^{2}/2k$, and entirely decoupled from ion vibrations. The
ion vibrates near a new equilibrium position, shifted by $f/k$, with
the ``old'' frequency $(k/M)^{1/2}$. As a result of the local ion
deformation, the total energy of the whole system decreases by
$E_{p}$ since a decrease of the electron energy by $-2E_{p}$
overruns an increase of the deformation energy $E_{p}$. Note that
the authors of Ref.~\cite{ratsw} made an illegitimate
replacement of the square of the occupation number operator $\hat{n}%
=c_{0}^{\dagger}c_{0}$ in Eq.~(\ref{eq:newH})
by its ``mean-field'' expression $\hat{n}^{2}=n_0\hat{%
n}$ which contains the average population of a single molecular level, $n_{0}
$, in disagreement with the exact identity, $\hat{n}^{2}=\hat{n}$. This
leads to a spurious \emph{self-interaction} of a single polaron with itself
[i.e. the term $\varepsilon=\varepsilon_{0}-n_{0}E_{p}$ instead of Eq.~(\ref
{eq:newLev})], and a resulting non-existent nonlinearity in the rate
equation.

Lattice deformation also strongly affects the interaction between electrons.
When a short-range deformation potential and molecular e-ph interactions are
taken into account together with the long-range Fr\"{o}hlich interaction,
they can overcome the Coulomb repulsion. The resulting interaction becomes
attractive at a short distance comparable to a lattice constant. The origin
of the attractive force between two small polarons can be readily understood
from a similar Holstein-like toy model as above \cite{alebook}, but with two
electrons on neighboring sites \textbf{1,2} interacting with an ion \textbf{3%
} between them, Fig.~\ref{fig:2sitepol}. For generality, we now assume that
the ion is a three-dimensional oscillator described by a displacement vector
$\mathbf{u}$, rather than by a single-component displacement $x$ as in
Eq.(1).

The vibration part of the Hamiltonian in the model is
\begin{equation}
H_{ph}=-{\frac{1}{{2M}}\frac{\partial^{2}}{{\partial\mathbf{u}}^{2}}}+{\frac{%
k\mathbf{u}^{2}}{{2}}},   \label{eq:Hv}
\end{equation}
Electron potential energies due to the Coulomb interaction with the ion are
about
\begin{equation}
V_{1,2}=V_{0}-\mathbf{u}\mathbf{\nabla}_{R_{1,2}}V_0(R_{1,2}),
\label{eq:V12}
\end{equation}
where $\mathbf{R_{1(2)}}$ is the vector connecting ion site
$\mathbf{3}$ with electron $\mathbf{1 (2)}$. Hence, the Hamiltonian
of the model is given by
\begin{equation}
H=E_{a}(\hat{n}_{1}+\hat{n}_{2})+\mathbf{u}\cdot(\mathbf{f_{1}}\hat{n}_{1}+%
\mathbf{f_{2}}\hat{n}_{2})-{\frac{1}{{2M}}\frac{\partial^{2}}{{\partial%
\mathbf{u}}^{2}}}+{\frac{k\mathbf{u}^{2}}{{2}}},
\end{equation}
where $\mathbf{f_{1,2}}=Ze^{2}\mathbf{e_{1,2}}/a^{2}$ is the Coulomb force,
and $\hat{n}_{1,2}$ are occupation number operators at every site. This
Hamiltonian is also readily diagonalized by the same displacement
transformation of the vibronic coordinate $\mathbf{u}$ as above,
\begin{equation}
\mathbf{u}=\mathbf{v}-\left( \mathbf{f_{1}}\hat{n}_{1}+\mathbf{f_{2}}\hat
{n}_{2}\right) /k.
\end{equation}
The transformed Hamiltonian has no electron-phonon coupling,
\begin{equation}
\tilde{H}=(\varepsilon_{0}-E_{p})(\hat{n}_{1}+\hat{n}_{2})+V_{ph}\hat{n}_{1}%
\hat{n}_{2}-{\frac{1}{{2M}}\frac{\partial^{2}}{{\partial\mathbf{v}}^{2}}}+{%
\frac{k\mathbf{v}^{2}}{{2}}},
\end{equation}
and it describes two small polarons at their atomic levels shifted by the
polaron level shift $E_{p}=f_{1,2}^{2}/2k$, which are entirely decoupled
from ion vibrations. As a result, the lattice deformation caused by two
electrons leads to an effective interaction between them, $V_{ph}$, which
should be added to their Coulomb repulsion, $V_{c}$,
\begin{equation}
V_{ph}=-\mathbf{f_{1}}\cdot\mathbf{f_{2}}/k.
\end{equation}
When $V_{ph}$ is negative and larger by magnitude than the positive $V_{c},$
the resulting interaction becomes attractive. That is $V_{ph}$ rather than $%
E_{p}$, which is responsible for the hysteretic behavior of MQDs, as
discussed below.


\subsection{ Exact solution}

The procedure, which fully accounts for all correlations in MQD is as
follows, see Ref.~\cite{AB03vibr}. The molecular Hamiltonian includes the
Coulomb repulsion, $U^{C}$, and the electron-vibron interaction as
\begin{align}
& H =\sum_{_{\mu}}\varepsilon_{_{\mu}}\hat{n}_{_{\mu}}+\frac{1}{2}%
\sum_{_{\mu }\neq\mu^{\prime}}U_{\mu\mu^{\prime}}^{C}\hat{n}_{_{\mu}}\hat{n}%
_{\mu^{\prime }}  \notag \\
& +\sum_{\mu,q}\hat{n}_{_{\mu}}\omega_{q}(\gamma_{\mu q}d_{q}+H.c.)+\sum
_{q}\omega_{q}(d_{q}^{\dagger}d_{q}+1/2).
\end{align}
Here $d_{q}$ annihilates phonons, $\omega_{q}$ is the phonon (vibron)
frequency, and $\gamma_{\mu q}$ are the e-ph coupling constant ($q$
enumerates the vibron modes). This Hamiltonian conserves the occupation
numbers of molecular states $\hat{n}_\mu$.

One can apply the canonical unitary transformation $e^{S}$, with
\begin{equation*}
S=-\sum_{q,\mu}\hat{n}_{\mu}\left( \gamma_{\mu q}d_{q}-H.c.\right)
\end{equation*}
integrating phonons out. The electron and phonon operators are transformed
as
\begin{equation}
\tilde{c}_{\mu}=c_{\mu}X_{\mu},\qquad X_{\mu}=\exp\left( \sum_{q}\gamma_{\mu
q}d_{q}-H.c.\right)
\end{equation}
and
\begin{equation}
\tilde{d}_{q}=d_{q}-\sum_{\mu}\hat{n}_{\mu}\gamma_{\mu q}^{\ast},
\end{equation}
respectively. This Lang-Firsov transformation shifts ions to new equilibrium
positions with no effect on the phonon frequencies. The diagonalization is
\emph{exact}:
\begin{equation}
\tilde{H}=\sum_{i}\tilde{\varepsilon}_{_{\mu}}\hat{n}_{\mu}+\sum_{q}\omega
_{q}(d_{q}^{\dagger}d_{q}+1/2)+{\frac{1}{{2}}}\sum_{\mu\neq\mu^{\prime}}U_{%
\mu\mu^{\prime}}\hat{n}_{\mu}\hat{n}_{\mu^{\prime}},  \label{eq:HLang}
\end{equation}
where
\begin{equation}
U_{\mu\mu^{\prime}}\equiv U_{\mu\mu^{\prime}}^{C}-2\sum_{q}\gamma_{\mu
q}^{\ast}\gamma_{\mu^{\prime}q}\omega_{q},   \label{eq:Umm1}
\end{equation}
is the renormalized interaction of polarons comprising their interaction via
molecular deformations (vibrons) and the original Coulomb repulsion, $%
U_{\mu\mu^{\prime}}^{C}$. The molecular energy levels are shifted by the
polaron level-shift due to the deformation created by the polaron,
\begin{equation}
\tilde{\varepsilon}_{_{\mu}}=\varepsilon_{_{\mu}}{-}\sum_{q}|\gamma_{\mu
q}|^{2}\omega_{q}.   \label{eq:eps}
\end{equation}

If we assume that the coupling to the leads is weak, so that the level width
$\Gamma \ll $ $|U|,$ we can find the current from \cite{meir92}
\begin{eqnarray}
I(V) &=&I_{0}\int_{-\infty }^{\infty }d\omega \left[ f_{1}(\omega
)-f_{2}(\omega )\right] \rho (\omega ),  \label{eq:Irho} \\
\rho (\omega ) &=&-\frac{1}{\pi }\sum_{\mu }\mathrm{Im}\hat{G}_{\mu
}^{R}(\omega ),
\end{eqnarray}
where $\left| \mu \right\rangle $ is a complete set of one-particle
molecular states, $f_{1(2)}(\omega)= \left( \exp\frac{\omega+\Delta \mp eV/2%
}{T}+1\right)^{-1}$ the Fermi function. Here $I_{0}=q\Gamma ,$ $\rho (\omega
)$ is the molecular DOS, $\hat{G}_{\mu }^{R}(\omega )$ is the Fourier
transform of the Green's function $\hat{G}_{\mu }^{R}(t)=$ $-i\theta
(t)\left\langle \left\{ c_{\mu }(t),c_{\mu }^{\dagger }\right\}
\right\rangle ,$ $\left\{ \cdots ,\cdots \right\} $ is the anticommutator, $%
c_{\mu }(t)=e^{iHt}c_{\mu }e^{-iHt},$ $\theta (t)=1$ for $t>0$ and zero
otherwise. We calculate $\rho (\omega )$ \emph{exactly} for the Hamiltonian (%
\ref{eq:HLang}), which includes both the Coulomb $U^{C}$ and e-ph
interactions.

The retarded GF becomes
\begin{eqnarray}
G_{\mu }^{R}(t) &=&-i\theta (t)[\left\langle c_{\mu }(t)c_{\mu }^{\dagger
}\right\rangle \left\langle X_{\mu }(t)X_{\mu }^{\dagger }\right\rangle
\notag \\
&&+\left\langle c_{\mu }^{\dagger }c_{\mu }(t)\right\rangle \left\langle
X_{\mu }^{\dagger }X_{\mu }(t)\right\rangle ].
\end{eqnarray}
The phonon correlator is simply
\begin{eqnarray}
\left\langle X_{\mu }(t)X_{\mu }^{\dagger }\right\rangle &=&\exp \sum_{q}%
\frac{|\gamma _{\mu q}|^{2}}{\sinh \frac{\beta \omega _{q}}{2}}  \notag \\
&&\times \left[ \cos \left( \omega t+i\frac{\beta \omega _{q}}{2}\right)
-\cosh \frac{\beta \omega _{q}}{2}\right] ,  \label{eq:Xcorr}
\end{eqnarray}
where the inverse temperature $\beta =1/T$, and $\left\langle X_{\mu
}^{\dagger }X_{\mu }(t)\right\rangle =\left\langle X_{\mu }(t)X_{\mu
}^{\dagger }\right\rangle ^{\ast }.$ The remaining GFs $\left\langle c_{\mu
}(t)c_{\mu }^{\dagger }\right\rangle $, are found from the equations of
motion \emph{exactly}. Finally, for the simplest case of a coupling to a
single mode with the characteristic frequency $\omega _{0}$ and $\gamma
_{q}\equiv \gamma $ we find \cite{AB03vibr}
\begin{eqnarray}
G_{\mu }^{R}(\omega ) &=&\mathcal{Z}\sum_{r=0}^{d-1}C_{r}(n)\sum_{l=0}^{%
\infty }I_{l}\left( \xi \right)  \notag \\
&&\biggl[e^{\frac{\beta \omega _{0}l}{2}}\left( \frac{1-n}{\omega
-rU-l\omega _{0}+i\delta }+\frac{n}{\omega -rU+l\omega _{0}+i\delta }\right)
\notag \\
&&+(1-\delta _{l0})e^{-\frac{\beta \omega _{0}l}{2}}  \notag \\
&&\times \biggl(\frac{1-n}{\omega -rU+l\omega _{0}+i\delta }+\frac{n}{\omega
-rU-l\omega _{0}+i\delta }\biggr )\biggr],  \label{eq:GwT}
\end{eqnarray}
where
\begin{equation}
\mathcal{Z}=\exp \left( -\sum_{\mathbf{q}}|\gamma _{q}|^{2}\coth \frac{\beta
\omega _{q}}{2}\right)
\end{equation}
is the familiar \emph{polaron narrowing} factor, the degeneracy factor
\begin{equation}
C_{r}(n)=\frac{(d-1)!}{r!(d-1-r)!}n^{r}(1-n)^{d-1-r},
\end{equation}
$\xi =|\gamma |^{2}/\sinh \frac{\beta \omega _{0}}{2},$ $I_{l}\left( \xi
\right) $ the modified Bessel function, and $\delta _{lk}$ the Kroneker
symbol. The important feature of the DOS, Eq.~(\ref{eq:Irho}), is its
nonlinear dependence on the occupation number $n.$ It contains full
information about all possible correlation and inelastic effects in
transport, in particular, all the phonon sidebands.

To simplify our discussion, we assume that the Coulomb integrals do not
depend on the orbital index, i.e. $U_{\mu\mu^{\prime}}=U$, and consider a
coupling to a single vibrational mode, $\omega_{q}=\omega_{0}$. Applying the
same transformation to the retarded Green's function, one obtains the exact
spectral function \cite{AB03vibr} for a $d-\mathrm{fold}$ degenerate MQD
(i.e. the density of molecular states, DOS) as
\begin{align}
& \rho(\omega)=\mathcal{Z}d\sum_{r=0}^{d-1}C_{r}(n)\sum_{l=0}^{\infty}I_{l}%
\left( \xi\right)  \notag \\
& \times\biggl[e^{\beta\omega_{0}l/2}\left[ (1-n)\delta(\omega-rU-l\omega
_{0})+n\delta(\omega-rU+l\omega_{0})\right]  \notag \\
& +(1-\delta_{l0})e^{-\beta\omega_{0}l/2}[n\delta(\omega-rU-l\omega _{0})
\notag \\
& +(1-n)\delta(\omega-rU+l\omega_{0})]\biggr],   \label{eq:rho}
\end{align}
where $\mathcal{Z}=\exp\left[ -|\gamma|^2\coth\frac{\beta\omega_{0}}{2}%
\right] $, is the above polaron narrowing factor, $\xi=|\gamma|^{2}/\sinh(%
\beta\omega_{0}/2),$ $\beta=1/T$, and $\delta_{lk}$ the Kroneker symbol.

The important feature of DOS, Eq.~(\ref{eq:rho}), is its nonlinear
dependence on the average electronic population $n=\left\langle
c_{\mu}^{\dagger}c_{\mu }\right\rangle ,$ which leads to the switching,
hysteresis, and other nonlinear effects in I-V characteristics for $d>2$
\cite{AB03vibr}. It appears due to \emph{correlations} between \emph{%
different} electronic states via the correlation coefficients $C_{r}(n)$.
There is no nonlinearity if the dot is nondegenerate, $d=1,$ since $%
C_{0}(n)=1$. In this simple case the DOS, Eq.~(\ref{eq:rho}), is a \emph{%
linear} function of the average population that can be found as a textbook
example of an exactly solvable problems \cite{mahan}.


In the present case of MQD weakly coupled with leads, one can apply the
Fermi-Dirac golden rule to obtain an equation for $n.$ Equating incoming and
outgoing numbers of electrons in MQD per unit time we obtain the
self-consistent equation for the level occupation $n$ as
\begin{eqnarray}
&&(1-n)\int_{-\infty }^{\infty }d\omega \left\{ \Gamma _{1}f_{1}(\omega
)+\Gamma _{2}f_{2}(\omega )\right\} \rho (\omega )  \notag \\
&=&n\int_{-\infty }^{\infty }d\omega \left\{ \Gamma _{1}[1-f_{1}(\omega
)]+\Gamma _{2}[1-f_{2}(\omega )]\right\} \rho (\omega ),  \label{eq:neq}
\end{eqnarray}
where $\Gamma _{1(2)}$ are the transition rates from left (right) leads to
MQD, and $\rho (\omega )$ is found from Eqs.~(\ref{eq:GwT}) and (\ref
{eq:Irho}). For $d=1,2$ the kinetic equation for $n$ is linear, and the
switching is \emph{absent.} Switching appears for $d\geq 3,$ when the
kinetic equation becomes non-linear. Taking into account that $\int_{-\infty
}^{\infty }\rho (\omega )=d$, Eq.~(\ref{eq:neq}) for the symmetric leads, $%
\Gamma _{1}=\Gamma _{2},$ reduces to
\begin{equation}
2nd=\int d\omega \rho \left( \omega \right) \left( f_{1}+f_{2}\right) ,
\end{equation}
which automatically satisfies $0\leq n \leq 1$. Explicitly, the
self-consistent equation for the occupation number is
\begin{equation}
n=\frac{1}{2}\sum_{r=0}^{d-1}C_{r}(n)[na_{r}^+ + (1-n)b_{r}^+],
\label{eq:neq1}
\end{equation}
where
\begin{eqnarray}
a_{r}^+ &=&\mathcal{Z}\sum_{l=0}^{\infty }I_{l}\left( \xi \right) \biggr(e^{%
\frac{\beta \omega _{0}l}{2}}[f_{1}(rU-l\omega _{0})+f_{2}(rU-l\omega _{0})]
\notag \\
&&+(1-\delta _{l0})e^{-\frac{\beta \omega _{0}l}{2}}[f_{1}(rU+l\omega
_{0})+f_{2}(rU+l\omega _{0})]\biggr),  \label{eq:a} \\
b_{r}^+ &=&\mathcal{Z}\sum_{l=0}^{\infty }I_{l}\left( \xi \right) \biggr(e^{%
\frac{\beta \omega _{0}l}{2}}[f_{1}(rU+l\omega _{0})+f_{2}(rU+l\omega _{0})]
\notag \\
&&+(1-\delta _{l0})e^{-\frac{\beta \omega _{0}l}{2}}[f_{1}(rU-l\omega
_{0})+f_{2}(rU-l\omega _{0})]\biggr).  \label{eq:b}
\end{eqnarray}
The current is expressed as
\begin{equation}
j\equiv \frac{I(V)}{dI_{0}}=\sum_{r=0}^{d-1}Z_{r}(n)[na_{r}^{-
}+(1-n)b_{r}^{-}],
\end{equation}
where
\begin{eqnarray}
a_{r}^{-} &=&\mathcal{Z}\sum_{l=0}^{\infty }I_{l}\left( \xi \right) \biggr(%
e^{\frac{\beta \omega _{0}l}{2}}[f_{1}(rU-l\omega _{0})-f_{2}(rU-l\omega
_{0})]  \notag \\
&&+(1-\delta _{l0})e^{-\frac{\beta \omega _{0}l}{2}}[f_{1}(rU+l\omega
_{0})-f_{2}(rU+l\omega _{0})]\biggr),  \label{eq:at} \\
b_{r}^{- } &=&\mathcal{Z}\sum_{l=0}^{\infty }I_{l}\left( \xi \right) \biggr(%
e^{\frac{\beta \omega _{0}l}{2}}[f_{1}(rU+l\omega _{0})-f_{2}(rU+l\omega
_{0})]  \notag \\
&&(1-\delta _{l0})e^{-\frac{\beta \omega _{0}l}{2}}[f_{1}(rU-l\omega
_{0})-f_{2}(rU-l\omega _{0})]\biggr).  \label{eq:bt}
\end{eqnarray}

\subsection{ Absence of switching of single- or double-degenerate MQD}

\begin{figure}[ptb]
\begin{center}
\includegraphics[angle=0, width=1\textwidth]
{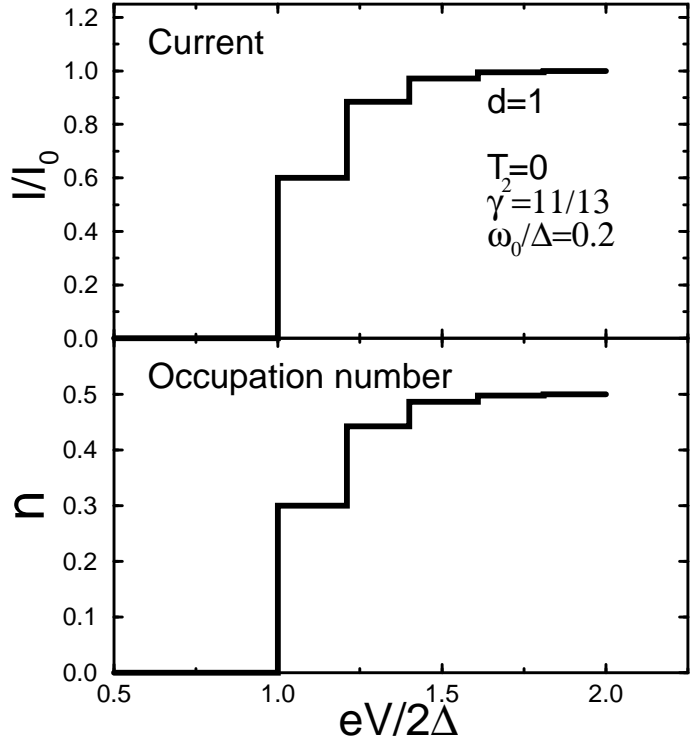}
\end{center}
\caption{Current-voltage characteristic of the nondegenerate ($d=1)$ MQD at $%
T=0,$ $\protect\omega_{0}/\Delta=0.2$, and $\protect\gamma^{2}=11/13$. There
is the phonon ladder in I-V , but no hysteresis.}
\label{fig:deg1}
\end{figure}
\begin{figure}[ptb]
\begin{center}
\includegraphics[angle=-0, width=1\textwidth]
{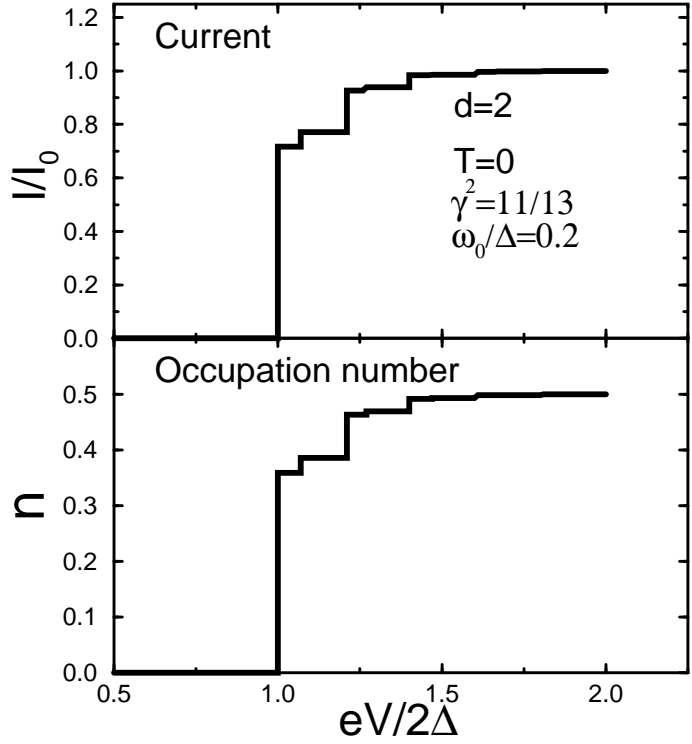}
\end{center}
\caption{Current-voltage characteristic of two-fold degenerate MQDs ($d=2)$
does not show hysteretic behavior. The parameters are the same as in Fig.~%
\ref{fig:deg1}. Larger number of elementary processes for conductance
compared to the nondegenerate case of $d=1$ generates more steps in the
phonon ladder in comparison with Fig.~\ref{fig:deg1}. }
\label{fig:deg2}
\end{figure}
%

If the transition rates from electrodes to MQD are small, $\Gamma \ll \omega
_{0}$, the rate equation for $n$ and the current, $I(V)$ are readily
obtained by using the exact molecular DOS, Eq.~(\ref{eq:rho}) and the
Fermi-Dirac Golden rule. In particular, for the nondegenerate MQD and $T=0$K
the result is
\begin{equation}
n=\frac{b_{0}^{+}}{2+b_{0}^{+}-a_{0}^{+}},
\end{equation}
and
\begin{equation}
j=\frac{2b_{0}^{-}+a_{0}^{-}b_{0}^{+}-a_{0}^{+}b_{0}^{-}}{%
2+b_{0}^{+}-a_{0}^{+}}.
\end{equation}
The general expressions for the coefficients are given at arbitrary
temperatures in Ref.\cite{AB03vibr}, and they are especially simple in
low-temperature limit:
\begin{align}
& a_{0}^{\pm }=\mathcal{Z}\sum_{l=0}^{\infty }\frac{|\gamma |^{2l}}{l!}%
[\theta (l\omega _{0}-\Delta +eV/2)  \notag \\
& \pm \theta (l\omega _{0}-\Delta -eV/2)], \\
& b_{0}^{\pm }=\mathcal{Z}\sum_{l=0}^{\infty }\frac{|\gamma |^{2l}}{l!}%
[\theta (-l\omega _{0}-\Delta +eV/2)  \notag \\
& \pm \theta (-l\omega _{0}-\Delta -eV/2)],
\end{align}
$j=I/I_{0}$, $I_{0}=ed\Gamma $, $\Delta $ is the position of the MQD level
with respect to the Fermi level at $V=0$, and $\theta (x)=1$ if $x>0$ and
zero otherwise. The current is single valued, Fig.~\ref{fig:deg1}, with the
familiar steps due to phonon-side bands. The double-degenerate level
provides more elementary processes for conductance reflected in larger
number of steps on phonon ladder compared to $d=2$ case, Fig.~\ref{fig:deg2}.

On the contrary, the mean-field approximation (MFA) leads to the
opposite conclusion. Galperin \textit{et al}.~\cite{ratsw} have
replaced the occupation number operator $\hat{n}$ in the e-ph
interaction by the average population $n_{0}$ [Eq.~(2) of
Ref.~\cite{ratsw}] and found the average steady-state vibronic
displacement $\langle d+d^{\dagger}\rangle$ proportional to $n_{0} $
(this is an explicit \emph{neglect} of all quantum fluctuations on
the dot accounted for in the exact solution). Then, replacing the
displacement operator $d+d^{\dagger}$ in the bare Hamiltonian,
Eq.~(11), by its average, Ref.~\cite{ratsw}, they obtained a new
molecular level,
$\tilde{\varepsilon}_{0}=\varepsilon_{0}-2\varepsilon_{reorg}n_{0}$
shifted linearly with the average population of the level. This is
in stark disagreement with the conventional constant polaronic level
shift, Eq.~(\ref {eq:newLev},\ref{eq:eps}) ($\varepsilon_{reorg}$ is
$|\gamma|^{2}\omega_{0}$ in our notations). The MFA spectral
function turned out to be highly nonlinear as a function of the
population, e.g. for the weak-coupling with
the leads $\rho(\omega)=\delta(\omega-\varepsilon_{0}-2%
\varepsilon_{reorg}n_{0}),$ see Eq.~(17) in Ref.~\cite{ratsw}. As a result,
the authors of Ref.\cite{ratsw} have found multiple solutions for the
steady-state population, Eq.~(15) and Fig.~1, and switching, Fig.~4 of Ref.~
\cite{ratsw}, which actually do not exist being an artefact of the model.

In the case of a double-degenerate MQD, $d=2,$ there are two terms, which
contribute to the sum over $r$, with $C_{0}(n)=1-n$ and $C_{1}(n)=n.$ The
rate equation becomes a quadratic one \cite{AB03vibr}. Nevertheless there is
only one physical root for any temperature and voltage, and the current is
also single-valued, Fig.~3.

Note that the mean-field solution by Galperin \textit{et al.} \cite{ratsw}
applies at any ratio $\Gamma/\omega_{0},$\ including the limit of interest
to us, $\Gamma\ll\omega_{0}.$ where their transition between the states with
$n_{0}=0$ and $1$ only sharpens, but none of the results change. Therefore,
MFA predicts a current bistability in the system where it does not exist at $%
d=1.$ Ref. \cite{ratsw} plots the results for $\Gamma\geq\omega_{0},$ $%
\Gamma\approx0.1-0.3$ eV, which corresponds to molecular bridges
with a resistance of about a few $100$~k$\Omega.$ Such model
``molecules'' are rather ``metallic'' in their conductance and could
hardly show any bistability at all because carriers do not have time
to interact with vibrons on the molecule. Indeed, taking into
account the coupling with the leads beyond the second order and the
coupling between the molecular and bath phonons could hardly provide
any non-linearity because these couplings do not depend on the
electron population. This rather obvious conclusion for molecules
strongly coupled to the electrodes can be reached in many ways, see
e.g. a derivation in Refs.~\cite{millis05,mozyr06}. While Refs.
\cite {millis05,mozyr06} do talk about telegraph current noise in
the model, there is no hysteresis in the adiabatic regime,
$\Gamma\gg\omega_{0}$ either. This result certainly has nothing to
do with our mechanism of switching \cite {AB03vibr} that applies to
molecular quantum dots ($\Gamma\ll\omega_{0})$ with $d>2.$ Such
regime has not been studied in Refs.~\cite
{millis05,mozyr06,millis04}, which have applied the adiabatic
approximation, as being ``too challenging problem''. Nevertheless,
Mitra \textit{et al}. \cite{millis04} have misrepresented our
formalism \cite{AB03vibr} claiming that it ``lacks of
renormanlization of the dot-lead coupling'' (due to electron-vibron
interaction), or ``treats it in an average manner''. In fact, the
formalism \cite{AB03vibr} is exact, fully taking into account the
polaronic renormalization, phonon-side bands and polaron-polaron
correlations in the exact molecular DOS, Eq.~(\ref{eq:rho}).

As a matter of fact, most of the molecules are very resistive, so the actual
molecular quantum dots are in the regime we study, see Ref.\cite{mqdexp}.
For example, the resistance of fully conjugated three-phenyl ring Tour-Reed
molecules chemically bonded to metallic Au electrodes \cite{ReedNDR} exceeds
$1$G$\Omega$. Therefore, most of the molecules of interest to us are in the
regime that we discussed, not that of Refs.\cite{millis05,mozyr06}.

\subsection{Nonlinear rate equation and switching}

\begin{figure}[ptb]
\begin{center}
\includegraphics [angle=0, width=1\textwidth]
{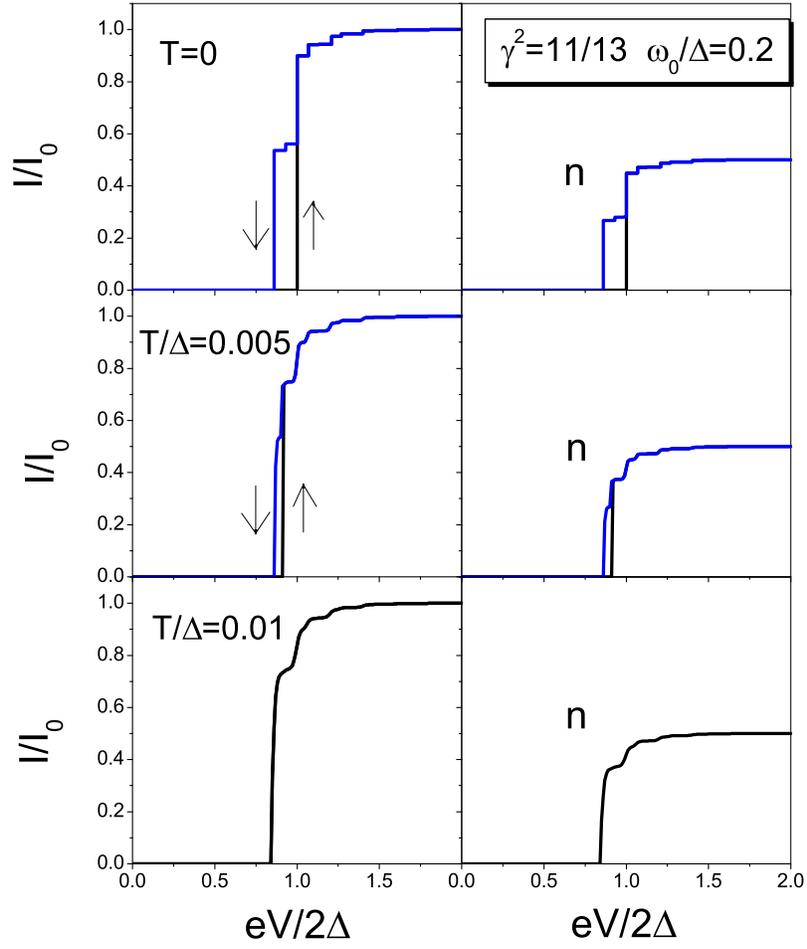}
\end{center}
\caption{ The I-V curves for tunneling through the molecular quantum dot,
Fig.~\ref{fig:moldot}b with the electron-vibron coupling constant $\protect%
\gamma ^{2}=11/13$. }
\label{fig:g1113}
\end{figure}

Now, consider the case $d=4$, where the rate equation is non-linear, to see
if it produces multiple physical solutions. For instance, in the limit $%
|\gamma |\ll 1,$ $T=0,$ where we have $b_{r}=a_{r}$, $\mathcal{Z}=1,$ the
remaining interaction is $U=U^{C}<0$, we recover the negative-$U$ model \cite
{alebrawil}, and the kinetic equation for $d=4$ is
\begin{equation}
2n=1-(1-n)^{3}  \label{eq:ncub}
\end{equation}
in the voltage range $\Delta -|U|<eV/2<\Delta $. This equation has
an additional nontrivial physical solution $n=(3-\sqrt{5})/2=0.38$.
The current is simplified as $I/I_{0}=2n.$ The current-voltage
characteristics will show a \emph{hysteretic} behavior in this case
for $d=4. $ When the voltage increases from zero, 4-fold degenerate
MQD remains in a low-current state until the threshold
$eV_{2}/2=\Delta $ is reached. Remarkably, when the voltage
\emph{decreases} from the value above the threshold $V_{2}$, the
molecule remains in the high-current state down to the voltage
$eV_{1}/2=\Delta -|U|$ well below the threshold $V_{2}$.

In fact, there is a hysteresis of current at all values of the
electron-phonon constant $\gamma ,$ e.g. $\gamma ^{2}=11/13$
(selected in order to avoid an accidental commensurability of
$\omega _{0}$ and $U)$, Fig.~\ref{fig:g1113}. Indeed, the exact
equation for average occupation of the dot reads
\begin{eqnarray}
2n &=&(1-n)^{3}[na_{0}^{+}+(1-n)b_{0}^{+}]  \notag \\
&&+3n(1-n)^{2}[na_{1}^{+}+(1-n)b_{1}^{+}]  \notag \\
&&+3n^{2}(1-n)[na_{2}^{+}+(1-n)b_{2}^{+}]  \notag \\
&&+n^{3}[na_{3}^{+}+(1-n)b_{3}^{+}].
\end{eqnarray}
We solved this nonlinear equation for the case $\omega
_{0}/\Delta =0.2,$ $U^{C}=0,$ so that the attraction between electrons is $%
U=-2\gamma ^{2}\omega _{0}=-0.4$ (all energies are in units of
$\Delta $), and found an additional stable solution for an average
occupation number (and current) that results in a hysteresis curve,
Fig.~\ref{fig:g1113}. The bistability region shrinks down with
temperature, and the hysteresis loop practically closes at $T/\Delta
=0.01$. As we see from Eqs.~(\ref{eq:a}),(\ref{eq:b}), the electron
levels with phonon sidebands $\Delta\pm l\omega _{0,}$ $\Delta+U\pm
l\omega _{0,} $ $\Delta+2U\pm l\omega _{0,}$ $\Delta+3U\pm l\omega
_{0}$ with $l=0,1,...$ contribute to electron transport with
different weights, and this creates a complex picture of steps on
the I-V curve, Fig. \ref{fig:g1113}.

Note that switching required a degenerate MQD ($d>2)$ and the weak coupling
to the electrodes, $\Gamma \ll \omega _{0}.$ Different from the
non-degenerate dot, the rate equation for a multi-degenerate dot, $d>2$,
weakly coupled to the leads has multiple physical roots in a certain voltage
range and a hysteretic behavior due to \emph{correlations} between different
electronic states of MQD \cite{AB03vibr}.

Summarizing this Section, we have calculated the I-V characteristics of the
nondegenerate and two-fold degenerate MQDs showing no hysteretic behavior,
and conclude that mean field approximation \cite{ratsw} leads to a
non-existent hysteresis in a model that was solved exactly in Ref.~\cite
{AB03vibr}. Different from the non-degenerate and two-fold degenerate dots,
the rate equation for a multi-degenerate dot, $d>2$, weakly coupled to the
leads, has multiple physical roots in a certain voltage range showing
hysteretic behavior due to \emph{correlations} between different electronic
states of MQD \cite{AB03vibr}. Our conclusions are important for searching
of the current-controlled polaronic molecular switches. Incidentally, C$_{60}
$ molecules have the degeneracy $d=6$ of the lowest unoccupied level, which
makes them one of the most promising candidate systems, if the weak-coupling
with leads is secured.


\section{Role of defects in molecular transport}


Interesting behavior of electron transport in molecular systems, as
described above, refers to ideal systems without imperfection in ordering
and composition. In reality, one expects that there will be a considerable
disorder and defects in organic molecular films. As mentioned above, the
conduction through absorbed \cite{polymer78} and Langmuir-Blodgett \cite
{tredgold81} monolayers of fatty acids (CH$_{2})_{n}$ was associated with
\emph{defects}. An absence of tunneling through self-assembled monolayers of
C12-C18 (inferred from an absence of thickness dependence at room
temperature)\ has been reported by Boulas et al.~\cite{boulas96}. On the
other hand, the tunneling in alkanethiol SAMs was reported in \cite
{sakaguchi01,reed03}, with an exponential dependence of monolayer resistance
on the chain length $L$, $R_{\sigma }\propto \exp (\beta _{\sigma }L),$ and
no temperature dependence of the conductance in C8-C16 molecules was
observed over the temperatures $T=80-300$K \cite{reed03}.

The electrons in alkane molecules are tightly bound to the C atoms by $%
\sigma -$bonds, and the band gap (between the highest occupied molecular
orbital, HOMO, and lowest unoccupied molecular orbital, LUMO) is large, $%
\sim 9-10$eV \cite{boulas96}. In conjugated systems with $\pi -$electrons
the molecular orbitals are extended, and the HOMO-LUMO\ gap is
correspondingly smaller, as in e.g. polythiophenes, where the resistance was
also found to scale exponentially with the length of the chain, $R_{\pi
}\propto \exp (\beta _{\pi }L),$ with $\beta _{\pi }=0.35 \AA^{-1}$ instead
of $\beta _{\sigma }=1.08\AA^{-1}$\cite{sakaguchi01}. In stark contrast with
the temperature-independent tunneling results for SAMs \cite{reed03}, recent
extensive studies of electron transport through 2.8 nm thick eicosanoic acid
(C20) LB\ monolayers at temperatures 2K-300K have established that the
current is practically temperature independent below $T<60$K, but very
strongly temperature dependent at higher temperatures $T=60-300$K\cite
{DunTdep05}.

\begin{figure}[ptb]
\begin{center}
\includegraphics [angle=0, width=.75\textwidth]
{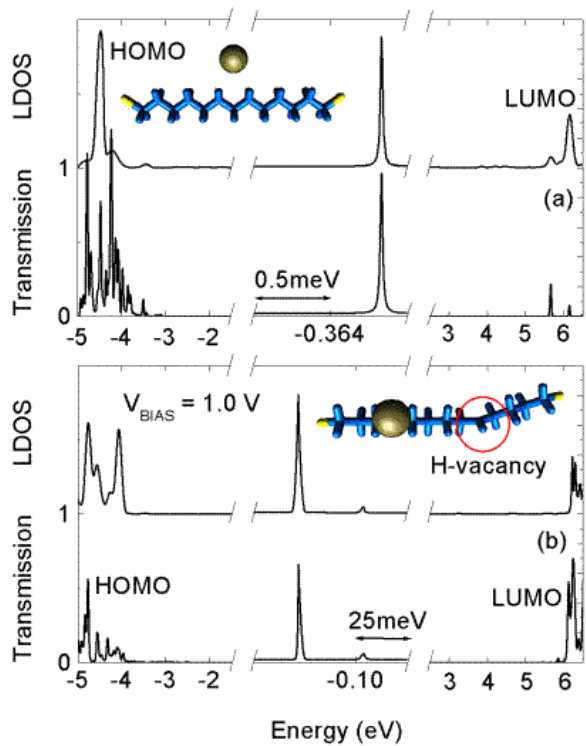}
\end{center}
\caption{ Local density of states and transmission as a function of energy
for (a) C13 with Au impurity and (b) C13 with Au impurity and H vacancy
(dangling bond). Middle sections show closeups of the resonant peaks due to
deep defect levels with respect to the HOMO and LUMO molecular states. The
HOMO-LUMO gap is about 10 eV. }
\label{fig:impTR}
\end{figure}

A large amount of effort went into characterizing the organic thin films and
possible defects there \cite{Ulman95,allara00,allaranoble04}. It has been
found that the electrode material, like gold, gets into the body of the
film, leading to the possibility of metal ions existing in the film as
single impurities and clusters. Electronic states on these impurity ions are
available for the resonant tunneling of carriers in very thin films (or
hopping in thicker films, a crossover between the regimes depending on the
thickness). Depending on the density of the impurity states, with increasing
film thickness the tunneling will be assisted by impurity ``chains'', with
an increasing number of equidistant impurities \cite{pollak73}. One-impurity
channels produce steps on the I-V curve but no temperature dependence,
whereas the inelastic tunneling through pairs of impurities at low
temperatures defines the temperature dependence of the film conductance, $%
G(T)\propto T^{4/3},$ and the voltage dependence of current $I(V)\propto
V^{7/3}$\cite{GlaMat88}. This behavior has been predicted theoretically and
observed experimentally for tunneling through amorphous Si \cite{Mac95} and
Al$_{2}$O$_{3}$\cite{Mood98}. Due to the inevitable disorder in a ``soft''
matrix, the resonant states on different impurities within a ``channel''
will be randomly moving in and out of resonance, creating mesoscopic
fluctuations of the I-V\ curve. The tunneling may be accompanied by
interaction with vibrons on the molecule, causing step-like features on the
I-V curve \cite{zhitvibr,AB03vibr}.

During processing, especially top electrode deposition, small clusters of
the electrode material may form in the organic film, causing Coulomb
blockade, which also can show up as steps on the I-V curve. It has long been
known that a strong applied field can cause localized damage to thin films,
presumably due to electromigration and the formation of conducting filaments
\cite{LBfila8671}. The damaged area was about 30nm in diameter in 40-160
monolayer thick LB\ films \cite{LBfila8671}(a) and 5-10$\mu $m in diameter
in films 500-5000\AA\ thick, and showed switching behavior under external
bias voltage cycling \cite{LBfila8671}(b). As discussed above, recent
spatial mapping of a conductance in LB\ monolayers of fatic acids with the
use of conducting AFM\ has revealed damage areas 30-100nm in diameter,
frequently appearing in samples after a ``soft'' electrical breakdown, which
is sometimes accompanied by a strong temperature dependence of the
conductance through the film\cite{Jennie04}.

\begin{figure}[ptb]
\begin{center}
\includegraphics [angle=0, width=.75\textwidth]
{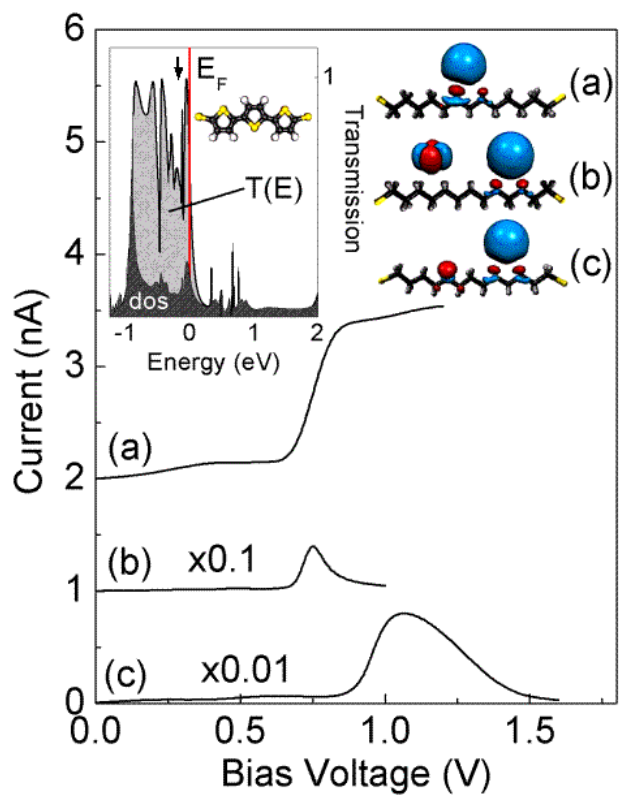}
\end{center}
\caption{ Current-voltage characteristics of an alkane chain C13 with (a)
single Au impurity (6s-state), (b) two Au impurities (5d and 6s-states on
left and right ions, respectively), and (c) Au impurity and H vacancy
(dangling bond). Double defects produce the negative differential resistance
peaks (b) and (c). Inset shows the density of states, transmission, and
stick model for polythiophene T3. There is significant transmission at the
Fermi level, suggesting an ohmic I-V characteristic for T3 connected to gold
electrodes. Disorder in the film may localize states close to the Fermi
level (schematically marked by arrow), which may assist in hole hopping
transport with an apparently very low activation energy (0.01-0.1eV), as is
observed.}
\label{fig:fig2iv}
\end{figure}

A crossover from tunneling at low temperatures to an activation-like
dependence at higher temperatures is expected for electron transport through
organic molecular films. There are recent reports about such a crossover in
individual molecules like the 2nm long Tour wire with a small activation
energy $E_{a}\approx 130$meV \cite{allaraBJTdep04}. Very small activation
energies on the order of 10-100 meV have been observed in polythiophene
monolayers \cite{Zhit04}. Our present results suggest that this may be a
result of interplay between the drastic renormalization of the electronic
structure of the molecule in contact with electrodes, and disorder in the
film (Fig.~\ref{fig:fig2iv}, right inset). We report the ab-initio
calculations of point-defect assisted tunneling through alkanedithiols S(CH$%
_{2})_{n}$S and thiophene T3 (three rings SC$_{4})$ self-assembled on gold
electrodes. The length of the alkane chain was in the range $n=9-15$.

We have studied single and double defects in the film: (i) single Au
impurity, Figs.~10a,11a, (ii) Au impurity and H vacancy (dangling bond) on
the chain, Figs.~\ref{fig:fig2iv}b, \ref{fig:impTR}c, (iii) a pair of Au
impurities, Fig.~\ref{fig:impTR}b, (iv) Au and a ``kink'' on the chain (one
C=C bond instead of a C-C bond). Single defect states result in steps on the
associated I-V curve, whereas molecules in the presence of two defects
generally exhibit a negative differential resistance (NDR). Both types of
behavior are generic and may be relevant to some observed unusual transport
characteristics of SAMs and LB\ films \cite
{polymer78,tredgold81,Jennie04,DunTdep05,allaraBJTdep04,Zhit04}. We have
used an ab-initio approach that combines the Keldysh non-equilibrium Green's
function (NEGF) method with self-consistent pseudopotential-based real space
density functional theory (DFT) for systems with \emph{open} boundary
conditions provided by semi-infinite electrodes under external bias voltage
\cite{Guo01,larade03}. All present structures have been relaxed with the
Gaussian98 code prior to transport calculations\cite{g98}. The conductance
of the system at a given energy is found from Eq.~(\ref{eq:gnegf}) and the
current from Eq.~(\ref{eq:cur}).

The equilibrium position of an Au impurity is about 3\AA\ away from the
alkane chain, which is a typical Van-der-Waals distance. As the density maps
show (Fig.~\ref{fig:fig2iv}), there is an appreciable hybridization between
the s- and d-states of Au and the sp- states of the carbohydrate chain.
Furthermore, the Au$^{+}$ ion produces a Coulomb center trapping a 6s
electron state at an energy $\epsilon _{i}=-0.35$eV with respect to the
Fermi level, almost in the middle of the HOMO-LUMO $\sim 10$eV\ gap in C${n}$%
. The tunneling evanescent resonant state is a superposition of the HOMO and
LUMO\ molecular orbitals. Those orbitals have a very complex spatial
structure, reflected in an asymmetric line shape for the transmission. Since
the impurity levels are very deep, they may be understood within the model
of \ ``short-range impurity potential'' \cite{LarMat87}. Indeed, the
impurity wave function \emph{outside} of the narrow well can be fairly
approximated as

\begin{equation}
\varphi (r)=\sqrt{\frac{2\pi }{\kappa }}\frac{e^{-\kappa r}}{r},
\label{eq:fii}
\end{equation}
where $\kappa $ is the inverse radius of the state, $\hbar ^{2}\kappa
^{2}/2m^{\ast }=E_{i},$ where $E_{i}=\Delta -\epsilon _{i}$ is the depth of
the impurity level with respect to the LUMO, and $\Delta $=LUMO$-F$ is the
distance between the LUMO\ and the Fermi level $F$ of gold and,
consequently, the radius of the impurity state $1/\kappa $ is small. The
energy distance $\Delta \approx 4.8$eV in alkane chains (CH$_{2})_{n}$\cite
{boulas96}\textbf{\ (}$\approx 5$eV from DFT calculations), and $m^{\ast
}\sim 0.4$ the effective tunneling mass in alkanes\cite{reed03}. For one
impurity in a rectangular tunnel barrier \cite{LarMat87} we obtain the
Breit-Wigner form of transmission $T(E,V)$,$\;$as before, Eq.~(\ref{eq:TEbw}%
). Using the model with the impurity state wave function (\ref{eq:fii}), we
may estimate for an Au impurity in C13 ($L=10.9$\AA ) the width $\Gamma
_{L}=\Gamma _{R}=1.2\times 10^{-6}$, which is within an order of magnitude
compared with the calculated value $1.85\times 10^{-5}$eV. The transmission
is maximal and equals unity when $E=\epsilon _{i}$ and $\Gamma _{L}=\Gamma
_{R},$ which corresponds to a symmetrical position for the impurity with
respect to the electrodes.

The electronic structure of the alkane backbone, through which the electron
tunnels to an electrode, shows up in the asymmetric lineshape, which is
substantially non-Lorentzian, Fig.~\ref{fig:impTR}. The current remains
small until the bias has aligned the impurity level with the Fermi level of
the electrodes, resulting in a step in the current, $I_{1}\approx \frac{2q}{%
\hbar }\Gamma _{0}e^{-\kappa L}$ (Fig.~\ref{fig:impTR}a). This step can be
observed only when the impurity level is not very far from the Fermi level $%
F,$ such that biasing the contact can produce alignment before a breakdown
of the device may occur. The most interesting situations that we have found
relate to the \textit{pairs} of point defects in the film. If the
concentration of defects is $c\ll 1$, the relative number of configurations
with pairs of impurities will be very small, $\propto c^{2}.$ However, they
give an exponentially larger contribution to the current. Indeed, the
optimal position of two impurities is symmetrical, a distance $L/2$ apart,
with current $I_{2}\propto e^{-\kappa L/2}.$ The conductance of a
two-impurity chain is \cite{LarMat87}
\begin{equation}
g_{12}(E)=\frac{4q^{2}}{\pi \hbar }\frac{\Gamma _{L}\Gamma _{R}t_{12}^{2}}{%
\left| (E-\epsilon _{1}+i\Gamma _{L})(E-\epsilon _{2}+i\Gamma
_{R})-t_{12}^{2}\right| ^{2}}.  \label{eq:g12}
\end{equation}
For a pair of impurities with slightly differing energies $%
t_{12}=2(E_{1}+E_{2})e^{-\kappa r_{12}}/\kappa r_{12},$ where $r_{12}$ is
the distance between them. The interpretation of the two-impurity channel
conductance (\ref{eq:g12})\ is fairly straightforward:\ if there were no
coupling to the electrodes, i.e. $\Gamma _{L}=\Gamma _{R}=0,$ the poles of $%
g_{12}$ would coincide with the bonding and antibonding levels of the
two-impurity ``molecule''. The coupling to the electrodes gives them a
finite width and produces, generally, two peaks in conductance, whose
relative positions in energy change with the bias. The same consideration is
valid for longer chains too, and gives an intuitive picture of the formation
of the impurity ``band''\ of states. The maximal conductivity $%
g_{12}=q^{2}/\pi \hbar $ occurs when $\epsilon _{1}=\epsilon _{2},$ $\Gamma
_{L}\Gamma _{R}=t_{12}^{2}=\Gamma _{2}^{2},$ where $\Gamma _{2}$ is the
width of the two-impurity resonance, and it corresponds to the symmetrical
position of the impurities along the normal to the contacts separated by a
distance equal to half of the molecule length, $r_{12}=L/2.$ The important
property of the two-impurity case is that it produces negative differential
resistance (NDR). Indeed, under external bias voltage the impurity levels
shift as
\begin{equation}
\epsilon _{i}=\epsilon _{i0}+qVz_{i}/L,  \label{eq:levels}
\end{equation}
where $z_{i}$ are the positions of the impurity atoms counted from the
center of the molecule. Due to disorder in the film, under bias voltage the
levels will be moving in and out of resonance, thus producing NDR\ peaks on
the I-V curve. The most pronounced negative differential resistance is
presented by a gold impurity next to a Cn chain with an H-vacancy on one
site, Fig.~\ref{fig:impTR}b (the defect corresponds to a dangling bond). The
defects result in two resonant peaks in transmission. Surprisingly, the H
vacancy (dangling bond) has an energy very close to the electrode Fermi
level $F$, with $\epsilon _{i}=-0.1$eV (Fig.~\ref{fig:impTR}b, right peak).
The relative positions of the resonant peaks move with an external bias and
cross at 1.2V, producing a pronounced NDR\ peak in the I-V curve, Fig.~\ref
{fig:fig2iv}c. No NDR peak is seen in the case of an Au impurity and a kink
C=C on the chain because the energy of the kink level is far from that of
the Au 6s impurity level. The calculated values of the peak current through
the molecules were large: $I_{p}\approx 90$ nA/molecule for an Au impurity
with H vacancy, and $\approx 5$ nA/molecule for double Au impurities.

We have observed a new mechanism for the NDR peak in a situation with two Au
impurities in the film. Namely, Au ions produce two sets of deep impurity
levels in Cn films, one stemming from the 6s orbital, another from the 5d
shell, as clearly seen in Fig.~\ref{fig:fig2iv}b (inset). The 5d-states are
separated in energy from 6s, so that now the tunneling through s-d pairs of
states is allowed in addition to s-s tunneling. Since the 5d-states are at a
lower energy than the s-state, the d- and s-states on different Au ions will
be aligned at a certain bias. Due to the different angular character of
those orbitals, the tunneling between the s-state on the first impurity and
a d-state on another impurity will be described by the hopping integral
analogous to the Slater-Koster $sd\sigma $ integral. The peak current in
that case is smaller than for the pair Au-H vacancy, where the overlap is of
$ss\sigma $ type (cf. Figs.~\ref{fig:fig2iv}b,c).

Thiophene molecules behave very differently since the $\pi -$states there
are conjugated and, consequently, the HOMO-LUMO\ gap is much narrower, just
below 2 eV. The tail of the HOMO\ state in the T3 molecule (with three
rings) has a significant presence near the electrode Fermi level, resulting
in a practically ``metallic'' density of states and hence ohmic I-V\
characteristic. This behavior is quite robust and is in apparent
disagreement with experiment, where tunneling has been observed\cite
{sakaguchi01}. However, in actual thiophene devices the contact between the
molecule and electrodes is obviously very poor, and it may lead to unusual
current paths and temperature dependence\cite{Zhit04}.

We have presented the first parameter-free DFT calculations of a class of
organic molecular chains incorporating single or double point defects. The
results suggest that the present generic defects produce deep impurity
levels in the film and cause a resonant tunneling of electrons through the
film, strongly dependent on the type of defects. Thus, a missing hydrogen
produces a level (dangling bond) with an energy very close to the Fermi
level of the gold electrodes $F.$ In the case of a single impurity, it
produces steps on the I-V curve when one electrode's Fermi level aligns with
the impurity level under a certain bias voltage. The two-defect case is much
richer, since in this case we generally see a formation of the
negative-differential resistance peaks. We found that the Au atom together
with the hydrogen vacancy (dangling bond) produces the most pronounced NDR
peak at a bias of 1.2V in C13. Other pairs of defects do not produce such
spectacular NDR peaks. A short range impurity potential model reproduces the
data very well, although the actual lineshape is different.

There is a remaining question of what may cause the strong temperature
dependence of conductance in ``simple'' organic films like [CH$_{2}]_{n}$.
The activation-like conductance $\propto \exp (-E_{a}/T)$ has been reported
with a small activation energy $E_{a}\sim 100-200$meV in alkanes \cite
{DunTdep05,allaraBJTdep04} and even smaller, 10-100 meV, in polythiophenes
\cite{Zhit04}. This is much smaller than the value calculated here for
alkanes and expected from electrical and optical measurements on C$_{n}$
molecules, $E_{a}\sim \Delta \sim 4$eV \cite{boulas96}, which correspond
nicely to the present results. In conjugated systems, however, there may be
rather natural explanation of small activation energies. Indeed, the HOMO in
T3 polythiophene on gold is dramatically broadened, shifted to higher
energies and has a considerable weight at the Fermi level. The upward shift
of the HOMO is just a consequence of the work function difference between
gold and the molecule. In the presence of (inevitable) disorder in the film
some of the electronic states on the molecules will be localized in the
vicinity of $E_{F}$. Those states will assist the thermally activated
hopping of holes within a range of small activation energies $\leq 0.1$~eV.
Similar behavior is expected for Tour wires\cite{allaraBJTdep04}, where $%
E_{F}-\mathrm{HOMO}\sim 1$~eV\cite{Guo01}(c), if the electrode-molecule
contact is poor, as is usually the case.

With regards to carrier hopping in monolayers of saturated molecules, one
may reasonably expect that in many studied cases the organic films are
riddled with metallic protrusions (filaments), emerging due to
electromigration in a very strong electric field, and/or metallic, hydroxyl,
etc. inclusions\cite{LBfila8671,Jennie04}. It may result in a much smaller
tunneling distance $d$ for the carriers and the image charge lowering of the
barrier. The image charge lowering of the barrier in a gap of width $d$ is $%
\Delta U=q^{2}\ln 4/(\epsilon d),$ meaning that a decrease of about 3.5eV
may only happen in an unrealistically narrow gap $d=2-3$\AA ~in a film with
dielectric constant $\epsilon =2.5,$ but it will add to the barrier
lowering. More detailed characterization and theoretical studies along these
lines may help to resolve this very unusual behavior. We note that such a
mechanism cannot explain the crossover with temperature from tunneling to
hopping reported for single molecular measurements, which has to be a
property of the device, but not a single molecule \cite{allaraBJTdep04}.

\section{Conclusions}

Studying molecules as possible building blocks for ultradense electronic
circuits is a fascinating quest that spans of over 30 years. It was inspired
decades ago by the notion that silicon technology is approaching its
limiting feature size, estimated at around 1985 to be about $1\mu $m\cite
{znanie85}. More than thirty years later and with FET gate lengths getting
below 10nm \cite{sub10nm}, the same notion that silicon needs to be replaced
at some point by other technologies floats again. We do not know whether
alternatives will continue to be steamrollered by silicon technology, which
is a leading nanotechnology at the moment, but the mounting resistance to
the famed Moore's law requires to look hard at other solutions for power
dissipation, leakage current, crosstalk, speed, and other very serious
problems. There are very interesting developments in studying electronic
transport through molecular films but the mechanisms of some observed
conductance ``switching'' and/or nonlinear electric behavior remain elusive,
and this interesting behavior remains intermittent and not very
reproducible. Most of the currently observed switching is extrinsic in
nature. For instance, we have discussed the effect of molecule-electrode
contact: the tilting of the angle at which the conjugated molecule attaches
to the electrode may dramatically change its conductance, and that probably
explains extrinsic ``telegraph'' switching observed in Tour wires \cite
{KB01,donhauser} and molecule reconfigurations may lead to similar phenomena
in other systems\cite{wolkow06}. Defects in molecular films have also been
discussed and may result in spurious peaks in I-V curves. We have outlined
some designs of the molecules that may demonstrate rectifying behavior,
which we call molecular ``quantum dots''. We have shown that at least in
some special cases molecular quantum dots may exhibit fast ($\sim $THz)
intrinsic switching.

The author is grateful to Jeanie Lau, Jason Pitters and Robert Wolkow for
kind permission to use their data and figures. The work has been partly
supported by DARPA.



%
%
%
%
%
%



\printindex

\end{document}